\RequirePackage{fix-cm}
\documentclass[twocolumn]{svjour3}          
\smartqed  
\usepackage{graphicx}
\usepackage{bm}
\usepackage[namelimits]{amsmath} 
\usepackage{amssymb}             
\usepackage{amsfonts}           
\usepackage{mathrsfs}            
\usepackage{amsmath}
\usepackage{booktabs}
\usepackage{stfloats}
\usepackage{graphicx}
\usepackage{algorithm}
\usepackage{algorithmicx}
\usepackage{algpseudocode}
\usepackage{listings}
\usepackage{changepage}
\usepackage{mathtools}
\usepackage{hyperref}
\usepackage{cite}
\usepackage{xcolor}
\usepackage{xurl}
\usepackage{cite}
\usepackage[caption=false,font=normalsize,labelfont=sf,textfont=sf]{subfig}

\usepackage{framed}

\def\eg{\emph{e.g.}}
\def\ie{\emph{i.e.}}
\newcommand{\norm}[1]{\left \lVert #1 \right \rVert}

\definecolor{formalshade}{rgb}{0.95,0.95,1}
\definecolor{darkblue}{rgb}{0.0, 0.0, 0.55}
\definecolor{tabblue}{RGB}{31, 119, 180}
\definecolor{tabred}{RGB}{214, 39, 40}

\hypersetup{
    colorlinks=true,
    linkcolor=tabblue,   
    citecolor=tabblue,
    urlcolor=tabblue
}

\newenvironment{formal}{%
  \MakeFramed{\advance\hsize-\width\FrameRestore}%
  \noindent\hspace{-4.55pt}
  \begin{adjustwidth}{}{7pt}%
  \vspace{2pt}\vspace{2pt}%
}
{%
  \vspace{2pt}\end{adjustwidth}\endMakeFramed%
}

\newcommand{\contribname}{Author Contributions}

\def\contribsection{\par\addvspace{17pt}\small\rmfamily
\trivlist\item[\hskip\labelsep{\bfseries\contribname}]}
\def\endcontribsection{\endtrivlist\addvspace{6pt}}

\newenvironment{auth}{\contribsection}{\endcontribsection}

\begin{document}
\title{Control of dynamical systems with neural networks
}


\author{
Lucas B\"ottcher
}


\institute{
Lucas B\"ottcher\at
Department of Computational Science and Philosophy, Frankfurt School of Finance and Management, Adickesallee 32--34, 60322, Frankfurt am Main, Germany and Laboratory for Systems Medicine, University of Florida, Gainesville, 32610-0225, Florida, United States of America\\
\email{l.boettcher@fs.de}
}

\date{\today}

\maketitle

\begin{abstract}
    Control problems frequently arise in scientific and industrial applications, where the objective is to steer a dynamical system from an initial state to a desired target state. Recent advances in deep learning and automatic differentiation have made applying these methods to control problems increasingly practical. In this paper, we examine the use of neural networks and modern machine-learning libraries to parameterize control inputs across discrete-time and continuous-time systems, as well as deterministic and stochastic dynamics. We highlight applications in multiple domains, including biology, engineering, physics, and medicine. For continuous-time dynamical systems, neural ordinary differential equations (neural ODEs) offer a useful approach to parameterizing control inputs. For discrete-time systems, we show how custom control-input parameterizations can be implemented and optimized using automatic-differentiation methods. Overall, the methods presented provide practical solutions for control tasks that are computationally demanding or analytically intractable, making them valuable for complex real-world applications.
\keywords{dynamical systems \and control \and neural networks \and neural ODEs \and reinforcement learning \and model predictive control \and conformal prediction}
\end{abstract}

\section{Introduction}
Control problems frequently arise in scientific and industrial applications, where the objective is to steer a dynamical system from an initial state to a desired target state. The theoretical foundations for addressing such problems are provided by control theory, which is historically connected to neuroscience and machine intelligence through the framework of cybernetics~\cite{rosenblueth1943behavior,wiener2019cybernetics} and approaches like connectionism~\cite{thorndike1932fundamentals,churchland1992computational,rumelhart1986pdp,rumelhart1986pdp2,miller1995neural}. More recently, advances in automatic differentiation~\cite{paszke2017automatic,jax2018github} and machine learning~\cite{lutter2019deep,zhong2019symplectic,jin2019pontryagin,asikis2022neural,bottcher2022near,bottcher2022ai,chee2022knode,bottcher2023gradient,mowlavi2023optimal,bottcher2023control,nghiem2023physics,chee2023uncertainty,mou2024model,chen2024accelerated,DELALEAU2025229,bottcher2025control,wang2025pidnodes} have further strengthened these connections. 

Modern machine-learning approaches are increasingly complementing traditional control methods and have shown potential in controlling complex dynamical systems, such as biomedical systems~\cite{raghu2017deep,wen2019online,bottcher2025control,steffen2025deep}, inventory systems~\cite{bottcher2023control,deng2023alibaba}, and fusion reactors~\cite{Degrave2022}.\footnote{Here, we use the term ``complex'' to refer to characteristics of a dynamical system that make its optimization and control challenging. These include factors such as high dimensionality of the underlying system, a large action space that defies efficient enumeration, and control inputs that are difficult to parameterize.} The challenge of applying standard control methods to complex dynamical systems was already recognized in 1971 by Alexey Ivakhnenko, a pioneer of deep learning~\cite{schmidhuber2015deep}, in his Polynomial Theory of Complex Systems~\cite{ivakhnenko1971polynomial}: 
\begin{formal}
Modern control theory, based on differential equations, is not an adequate tool for solving the problems of complex control systems. Constructing differential equations to trace input-output paths requires a deductive, deterministic approach. However, this approach is impractical for complex systems due to the difficulty of identifying these paths.
\vspace{1em}\\
\footnotesize{Alexey Ivakhnenko in Polynomial Theory of Complex Systems (1971)}
\end{formal}

Control methods involving neural networks have been applied to both discrete-time and continuous-time dynamical systems~\cite{lewis2020neural}. To maintain tractability in gradient-based parameter updates, early applications of neural network controllers (NNCs) often focused on shallow architectures and linear dynamics. When dealing with high-dimensional, nonlinear systems where direct gradient updates are intractable or computationally demanding, ``identifier'' neural networks have been used to approximate and replace the underlying system dynamics~\cite{lewis2020neural}. With the popularization of neural ordinary differential equations (neural ODEs) within the \texttt{PyTorch} framework~\cite{chen2018neural}, building on earlier contributions such as Runge--Kutta neural networks~\cite{wang1998runge}, the direct application of deep neural network architectures to high-dimensional, nonlinear dynamical systems has become more accessible.

Other common applications of neural networks in
control include neural Hamilton--Jacobi--Bellman (HJB) methods and deep reinforcement learning~\cite{abu2005nearly,PhysRevFluids.4.093902,de2023deep,gu2023optimal,wang2025energy}. Neural HJB methods have been applied to state-feedback control problems, relying on the existence (or approximability) of a smooth value function for the considered system~\cite{abu2005nearly}. Deep reinforcement learning is often used in model-free settings, where the system dynamics are unknown or non-differentiable~\cite{mizutani2004two,DBLP:conf/nips/JinABJ18,DBLP:conf/aaai/Yarats0KAPF21}. In this work, we focus on model-based methods, which directly incorporate a neural network into the known system dynamics. These approaches, sometimes referred to as actor-only reinforcement learning, can be more sample-efficient and converge more rapidly to near-optimal solutions than their model-free, actor-critic counterparts.

We examine how the expressive power of neural networks (\ie, their ability to approximate
a broad class of functions) can be leveraged to parameterize control inputs across a wide range of control and optimization problems. These problems span both discrete-time and continuous-time systems, as well as deterministic and stochastic dynamics. We will highlight applications across various domains, including biology, engineering, physics, and medicine. For continuous-time dynamical systems, neural ODEs provide a valuable approach to parameterizing control inputs. For discrete-time systems, we will describe how automatic differentiation can be used to implement and optimize custom control-input parameterizations.

This paper proceeds as follows. In Sections~\ref{sec:discrete_time} and~\ref{sec:cont_time}, we review selected neural-control methods for discrete- and continuous-time dynamics, complemented by new examples that illustrate key ideas and potential extensions. In Section~\ref{sec:comparison_mpc}, we compare a neural control approach with model predictive control (MPC) and discuss prior work on neural-MPC frameworks. In Section~\ref{sec:conformal_prediction}, we integrate conformal prediction into neural control systems to quantify uncertainty. We conclude by summarizing key results and listing relevant code repositories in Section~\ref{sec:conclusions}.
\begin{figure}
    \centering
    \includegraphics[width=0.48\textwidth]{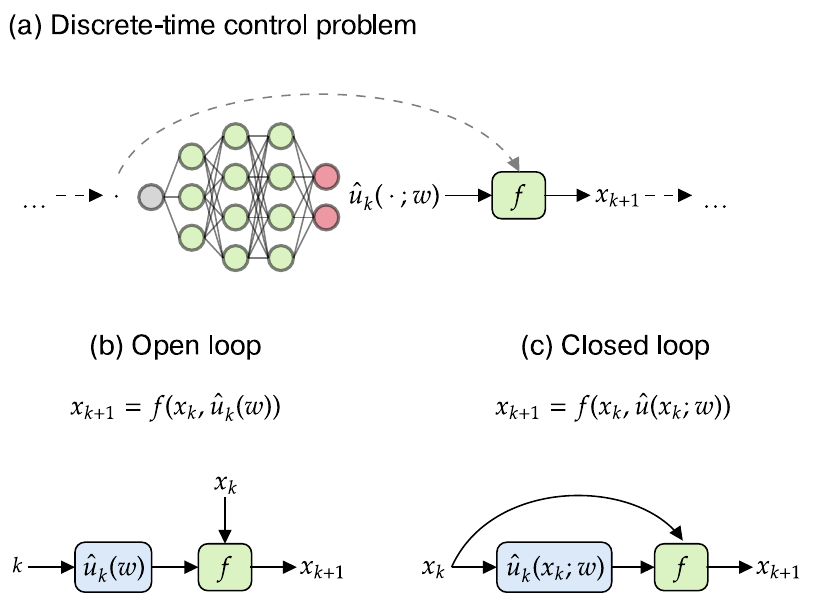}
    \caption{Schematic of a discrete-time control problem, where a control input $\hat{u}_k(\cdot; w)$ is parameterized by a neural network with parameters $w\in\mathbb{R}^N$. We refer to these control inputs as neural network controllers (NNCs). (a) The overall system structure, in which the NNC generates control inputs that influence the state transition function $f(\cdot)$. When the NNC depends on the system state $x_k$, a skip connection (dashed grey line) propagates state information directly from the controller input to $f(\cdot)$, creating a ResNet-like unfolding of the underlying dynamics. (b) An open-loop control scenario, where the control input $\hat{u}_k(w)$ depends on the time step $k$ but not on the system state $x_k$. (c) A closed-loop control scenario, where the control input $\hat{u}_k(x_k; w)$ is computed based on the current state $x_k$.}
    \label{fig:discrete_time_control}
\end{figure}
\section{Discrete-time dynamics}
\label{sec:discrete_time}
We first consider discrete-time dynamical systems of the form
\begin{equation}
    x_{k+1} = f(x_k, u_k)\,, \quad k\in\{0, \dots, T-1\}\,,
\end{equation}
where $x_k\in\mathbb{R}^n$ and $u_k\in\mathbb{R}^m$ represent the system state and control input at time step $k$, respectively, and $f\colon \mathbb{R}^n\times \mathbb{R}^m\rightarrow \mathbb{R}^n$ is the state transition function. We examine both open-loop and closed-loop control inputs, $\hat{u}_k(w)$ and $\hat{u}_k(x_k;w)$, parameterized by a neural network with parameters $w\in\mathbb{R}^N$. We refer to these control inputs as neural network controllers (NNCs), whose parameters are trained using standard optimizers such as Adam and RMSProp, as implemented in \texttt{PyTorch}.

In Fig.~\ref{fig:discrete_time_control}, we show a schematic of a discrete-time control problem with an NNC $\hat{u}_k(\cdot; w)$. When the control input depends on the system state $x_k$, a skip connection [dashed grey line in Fig.~\ref{fig:discrete_time_control}(a)] propagates state information directly from the controller input to the state transition function $f(\cdot)$, creating a ResNet-like unfolding of the underlying dynamics.\footnote{In a residual neural network (ResNet), each layer adds a residual $f(x)$ to its input $x$, so that the input to the next layer becomes $f(x)+x$. This is implemented via a skip connection that carries the input forward and adds it to the layer's output.} In Fig.~\ref{fig:discrete_time_control}(b), we illustrate an open-loop control scenario, where the control input $\hat{u}_k(w)$ depends only on the time step $k$. In contrast, Fig.~\ref{fig:discrete_time_control}(c) depicts a closed-loop control scenario, where the control input $\hat{u}_k(x_k;w)$ depends on the current state $x_k$, allowing the NNC to adapt its output based on the system’s behavior.
\begin{figure*}
    \centering
    \includegraphics{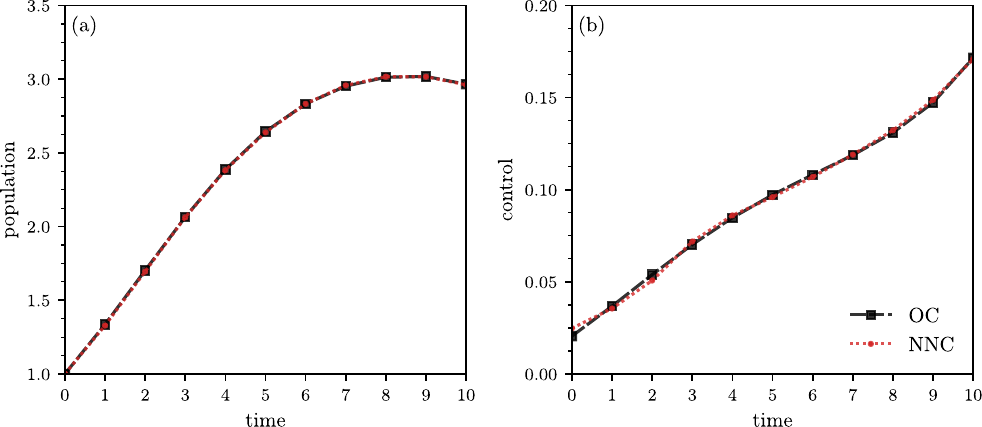}
    \caption{Comparison of control approaches for the Beverton--Holt model \eqref{eq:beverton_holt} with harvesting over a finite time horizon $T = 10$, using parameters $\gamma = 1$, $a = 0.1$, $c = 10$, and $r = 1.5$. The control objective is to maximize the total discounted net benefit over $T$, as defined in Eq.~\eqref{eq:bh_j1}. (a) Evolution of the population $x_k$ under the optimal control (OC) and the NNC approaches. (b) Corresponding control input over time. The population initially increases due to low control, then stabilizes as the control intensifies. The NNC closely resembles the OC solution.}
    \label{fig:beverton_holt}
\end{figure*}

To train NNCs, we define and optimize a suitable loss function that reflects the control objective. A common example is the finite-horizon cost functional
\begin{equation}
    J[\{x_k\}, \{u_k\}] = \sum_{k=0}^{T-1} L(x_k, u_k, k) + V(x_T)\,,
\label{eq:discrete_time_cost}
\end{equation}
where $L(x_k, u_k, k)$ denotes the running cost at time step $k$, and $V(x_T)$ is the terminal cost associated with the final state. Replacing $u_k$ with the parameterized control input $\hat{u}_k(\cdot; w)$, we optimize the cost functional $J[\{x_k\}, \{\hat{u}_k\}]$ using automatic differentiation to update the NNC parameters $w$.

We now present several examples to illustrate the applicability of NNCs to discrete-time dynamical systems. The first is a simple, illustrative example based on a deterministic discrete-time system describing the evolution of a single-species population under harvesting~\cite{beverton1957dynamics} using an open-loop controller. The remaining two examples, which focus on predator-prey interactions~\cite{bottcher2025control} and inventory dynamics~\cite{bottcher2023control}, involve stochasticity and closed-loop controllers.
\subsection{Illustrative example}
\label{sec:discrete_time_example}
As a basic illustrative example, we consider a discrete-time optimal control problem describing the evolution of a single-species population subject to harvesting. The population dynamics follow a version of the Beverton--Holt model
\begin{equation}
    x_{k+1} = \frac{r x_k}{1 + a x_k}(1 - u_k)\,,
\label{eq:beverton_holt}
\end{equation}
where $x_k \geq 0$ denotes the population size at time $k\in\{0,\dots,T-1\}$, $r > 0$ is the intrinsic growth rate, $a > 0$ captures density-dependent effects (\eg, crowding or competition), and $u_k \in [0,1]$ represents the fraction of the population harvested at time $k$. The Beverton--Holt model was originally introduced by Beverton and Holt (1957) to describe the growth dynamics of fish populations~\cite{beverton1957dynamics}. Related ideas can be traced back to the early work of Baranov (1918)~\cite{baranov1918biological}, whose contributions continue to influence modeling approaches in fisheries science~\cite{kenchington2021baranov}.

The control objective is to maximize the total discounted net benefit over a finite time horizon $T$~\cite{whittle_discrete_eco_models}, given by
\begin{equation}
    J_1[\{x_k\}, \{u_k\}] = \sum_{k=0}^{T-1} \left[ \gamma^k x_k \frac{r x_k}{1 + a x_k} - c u_k^2 \right]\,,
    \label{eq:bh_j1}
\end{equation}
where $\gamma \geq 0$ is the discount factor and $c \geq 0$ penalizes large harvest rates, \eg, those arising from increasing marginal harvesting costs or ecological impacts.

We aim to determine a control sequence $\{u_k\}_{k=0}^{T-1}$ such that $u_k \in [0,1]$ for all $k$, in order to maximize $J_1[\{x_k\}, \{u_k\}]$ subject to the state dynamics~\eqref{eq:beverton_holt}. 

We use an NNC $\hat{u}_k(w)$ to parameterize the control input with parameters $w\in\mathbb{R}^N$, which we determine by minimizing $-J_1[\{x_k\}, \{\hat{u}_k\}]$. The NNC that we employ has five hidden layers with four rectified linear units (ReLUs) each.

As a baseline for the NNC approach, we also apply Pontryagin’s maximum principle to the Hamiltonian
\begin{align}
\begin{split}
    H=&\sum_{k=0}^{T-1} \left[\gamma^k u_k \frac{r x_k}{1 + a x_k} - c u_k^2 \right. \\
    &\quad \quad \left.+ \lambda_k \left(x_{k+1} - \frac{r x_k}{1 + a x_k}(1 - u_k)\right) \right]\,,
\end{split}
\end{align}
with the adjoint variables $\lambda_k$, as a necessary condition for obtaining the optimal control (OC) solution.

The corresponding control input satisfies
\begin{equation}
    u_k = \frac{\gamma^k + \lambda_k}{2c} \frac{r x_k}{1 + a x_k}
    \label{eq:bh_uk}
\end{equation}
and the adjoint variables evolve according to
\begin{equation}
    \lambda_{k-1} = \lambda_k \frac{r}{(1 + a x_k)^2}(1 - u_k) - \gamma^k u_k \frac{r}{(1 + a x_k)^2}\,.
    \label{eq:bv_lambdak}
\end{equation}

In Fig.~\ref{fig:beverton_holt}, we show the evolution of the population as well as the control trajectories obtained from the OC and NNC policies. The simulation is conducted over a finite time horizon of $T = 10$, with parameter values $\gamma = 1$, $a = 0.1$, $c = 10$, and $r = 1.5$. The population initially grows due to weak control and later stabilizes as the control signal increases. The NNC solution closely approximates the OC trajectory, while avoiding the iteration of the coupled control-adjoint system [see Eqs.~\eqref{eq:bh_uk} and \eqref{eq:bv_lambdak}].
\subsection{Predatory-prey dynamics}
\label{sec:predator_prey}
As a more involved example, we consider a predator-prey agent-based model (ABM) with three species $A$, $B$, and $C$~\cite{may1975nonlinear,pekalski1998three,wilensky1997netlogo,wilensky1999netlogo,bottcher2025control}. We denote the population sizes of species $A$, $B$, and $C$ at time step $k$ by $a_k$, $b_k$, and $c_k$, respectively. In an ecological context, this model can represent interactions such as those between grass, sheep, and wolves, or between plankton, forage fish, and predatory fish. Similar models also appear in systems biology. For example, in models of Aspergillosis, a common lung infection caused by the fungus \textit{Aspergillus}, the three species may correspond to iron (as a nutrient source), \textit{Aspergillus} (as prey), and macrophages (as predators)~\cite{oremland2016computational,ribeiro2022multi}. Generalized predator-prey models with even more species have found applications in studies of microbial communities~\cite{faust2012microbial}, whose continuous-time description will be the subject of Section~\ref{sec:predator_prey_cont}.
\begin{figure}
    \centering
    \includegraphics[width=0.49\textwidth]{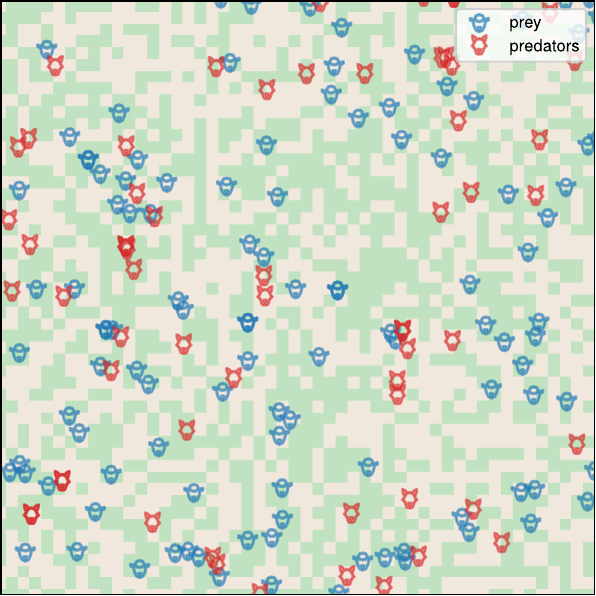}
    \caption{Predator-prey ABM. Snapshot of a three-species predator-prey ABM simulation on a $51\times 51$ grid. Grid cells colored green and light brown indicate nutrient-rich and nutrient-poor regions, respectively.}
    \label{fig:abm}
\end{figure}

We simulate the three-species predator-prey dynamics on an $L \times L$ periodic grid using an ABM defined by the following rules. Each grid cell can be in one of two states: (i) nutrient-rich or (ii) nutrient-poor. Prey move randomly with a directional bias toward the positive $x$-direction and rely on nutrient consumption to survive. The energy gain per unit of nutrient is $\kappa_1$. When a prey encounters a nearby nutrient-rich cell, it consumes the nutrients, causing the cell to switch to a nutrient-poor state. Nutrients in that cell regenerate after $\tau$ time steps.

Predators also perform a random walk with the same directional bias as the prey and consume prey when they occupy the same grid cell. The energy gained per consumed prey is $\kappa_2$. In each time step, both predators and prey lose one unit of energy to sustain their metabolism. Individuals die if their energy is smaller than 0. Predators and prey reproduce at rates $\alpha_1$ and $\alpha_2$, respectively.

In Fig.~\ref{fig:abm}, we show a snapshot of a three-species predator-prey ABM simulation on a $51 \times 51$ grid. Green and light brown cells indicate nutrient-rich and nutrient-poor regions, respectively. For additional details on this model, see \cite{wilensky1997netlogo}.

\begin{figure*}
    \centering
    \includegraphics{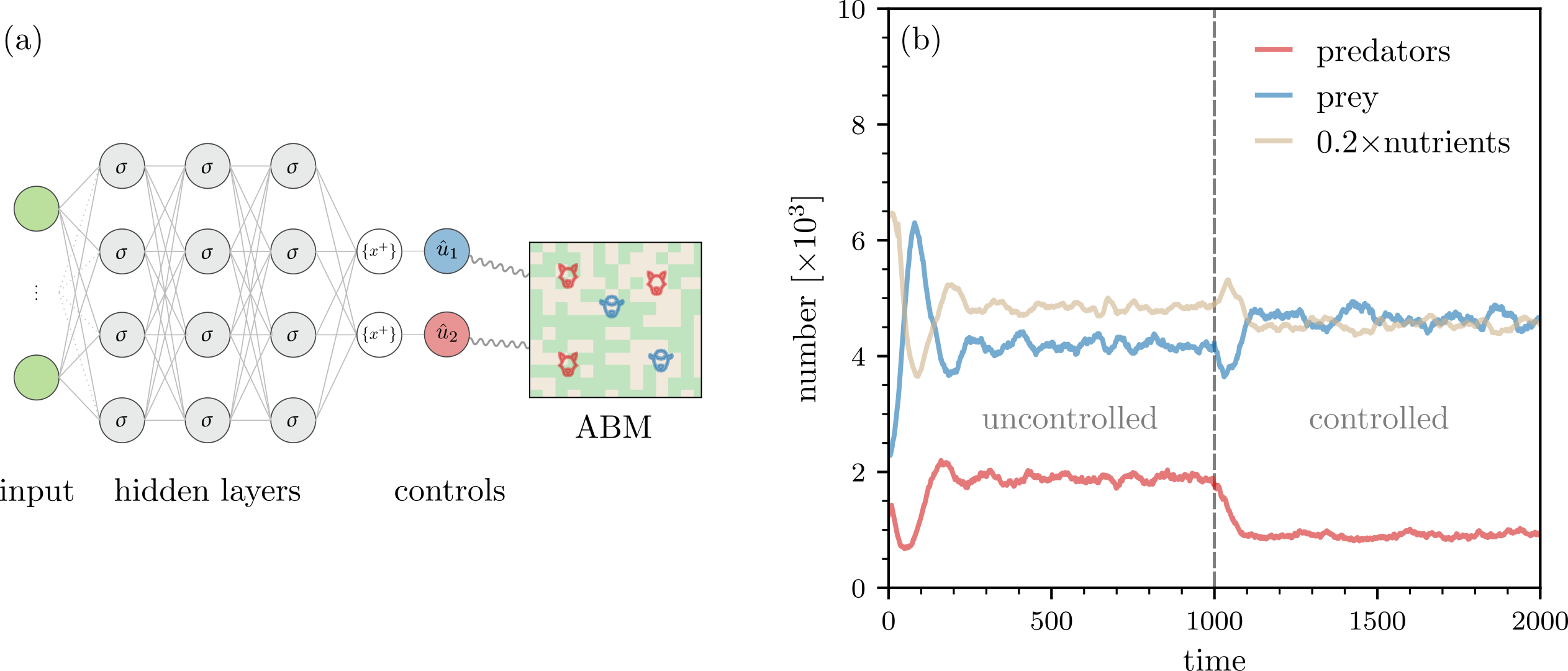}
    \caption{Control of a predator-prey ABM using an NNC. (a) To control the predator-prey ABM, we first define appropriate inputs and outputs for the NNC. Potential inputs include the population sizes $a_k$, $b_k$, and $c_k$ of species $A$, $B$, and $C$ at time step $k$. Since we aim to directly control predator and prey populations, the ANN produces two outputs, $\hat{u}_1$ and $\hat{u}_2$. A problem-specific straight-through estimator is used to obtain integer-valued control actions by removing the fractional part $\{\max\{0,\cdot\}\}$ from the positive hidden-layer outputs. We denote the hidden-layer activations by $\sigma$, and the straight-through estimator by $\{x^+\}$. (b) Evolution of predators, prey, and nutrient-right lattice sites  in a single realization of the predator-prey ABM. The vertical dashed gray line marks the time when the NNC is activated. The controller is designed to increase the mean number of prey by 10\% and reduce the mean number of predators by 50\%. Dashed blue and red lines indicate the target prey and predator levels, respectively (\ie, $\bar{b}^* = 4575$ and $\bar{c}^* = 948$). The simulation uses a $255 \times 255$ grid with initial conditions $b_0 = 2500$, $c_0 = 1250$, and parameters $\alpha_1 = 4.0$, $\alpha_2 = 5.0$, $\kappa_1 = 4.0$, $\kappa_2 = 20.0$, and $\tau = 30$~\cite{wilensky1997netlogo}. Initially, 50\% of the lattice sites are nutrient-rich.}
    \label{fig:predator_prey}
\end{figure*}

To control the predator-prey ABM, we define suitable inputs and outputs for an NNC. Potential inputs are the population sizes $a_k$, $b_k$, and $c_k$ of species $A$, $B$, and $C$ at time step $k$. As an example, we aim to shift the system toward a new steady state by directly adjusting the numbers of predators and prey. The NNC has two integer-valued outputs, $\hat{u}_1$ and $\hat{u}_2$ [see Fig.~\ref{fig:predator_prey}(a)]. If $\hat{u}_1 > 0$ ($\hat{u}_2 > 0$), a new prey (predator) is added at a randomly selected grid cell. If $\hat{u}_1 < 0$ ($\hat{u}_2 < 0$), a prey (predator) is removed from a randomly selected location. The underlying ABM used in our simulations consists of a $255 \times 255$ grid.

Before applying control, we allow the ABM to evolve freely for 1000 time steps to estimate steady-state population levels. For our chosen parameters, the mean numbers of predators and prey over the final 100 steps are approximately 1896 and 4159, respectively. We then run the controlled dynamics for an additional 1000 steps, giving a total time horizon $T = 2000$.

Our control objective is to increase the mean number of prey by 10\% and simultaneously reduce the mean number of predators by 50\% over the final $ T'$ time steps ($k \in \{T - T' + 1, \dots, T\}$) [see Fig.~\ref{fig:predator_prey}(b)]. Since such a substantial drop in predator numbers naturally leads to a rise in prey, the controller must suppress both populations relative to the original steady state. This goal can be achieved using a two-node NNC, with each output regulating the population of predators and prey, respectively. To train the NNC, we use the quadratic loss function
\begin{equation}
    J_2(w)=\big(\bar{b}(w)-\bar{b}^*\big)^2+\big(\bar{c}(w)-\bar{c}^*\big)^2\,,
\label{eq:loss}
\end{equation}
where $w \in \mathbb{R}^N$ are the parameters of the NNC, and $\bar{b}^*$ and $\bar{c}^*$ are the target values, corresponding to the desired mean numbers of prey and predators over the final $T'$ time steps. In this steady-state control example, we use $N = 2$ NNC parameters, $w = (w_1, w_2)^\top$, with target values $\bar{b}^* = 4575$, $\bar{c}^* = 948$, and $T' = 100$. The quantities
\begin{align}
    \bar{b}(w)=\frac{1}{T'}\sum_{k=1}^{T'} b_{(T-T'+k)}(w)
    \label{eq:reached_prey}
\end{align}
and
\begin{align}
    \bar{c}(w)=\frac{1}{T'}\sum_{k=1}^{T'} c_{(T-T'+k)}(w)
    \label{eq:reached_predators}
\end{align}
are the corresponding reached states. 

We parameterize the integer-valued control input $\hat{u}(b_k,c_k;w)$ according to
\begin{equation}
    \hat{u}(b_k,c_k;w) = 
    \begin{pmatrix}
    -(\max\{0,b_k w_1\}-\{\max\{0,b_k w_1\}\}) \\
    -(\max\{0,c_k w_2\}-\{\max\{0,c_k w_2\}\})
    \end{pmatrix}\,.
    \label{eq:control_two_node}
\end{equation}
Here, $\{x\}$ denotes the fractional part of $x$, defined as $\{x\} = x - \lfloor x \rfloor$ for $x > 0$, where $\lfloor \cdot \rfloor$ is the floor function. Training the NNC with the control input defined in Eq.~\eqref{eq:control_two_node} is based on a problem-specific straight-through estimator~\cite{714215,asikis2023multi,bottcher2023control,dyer2023}, which enables backpropagation with integer-valued outputs.

We use the two control inputs 
\begin{equation}
    \hat{u}_1(b_k;w_1)=-(\max\{0,b_k w_1\}-\{\max\{0,b_k w_1\}\})
\end{equation}
and 
\begin{equation}
    \hat{u}_2(c_k;w_2)=-(\max\{0,c_k w_2\}-\{\max\{0,c_k w_2\}\})
\end{equation}
to adjust the population sizes of prey and predators, respectively. These control inputs are set up such that they output negative integer-valued controls, meaning that a certain number of prey and predators will be removed from the ABM at each time step. For further details on the training of this NNC and an application of a multilayer NNC to transient control, see \cite{bottcher2025control}.

The lowest training loss, $J_1(w) \approx 74.09$, is achieved for parameters $w_1 = 0.0083$ and $w_2 = 0.0047$. The corresponding mean populations are approximately $\bar{b}(w) \approx 4573$ prey and $\bar{c}(w) \approx 956$ predators.

The learned NNC parameters $(0.0083,0.0047)$ are close to the optimal ones $(0.0083,0.0045)$, which have been determined by performing a grid search over the underlying parameter space~\cite{fonseca2025optimal}. To examine the uncertainty in the target quantities (\ie, the numbers of prey and predators), we tested the steady-state NNC on 50 previously unseen ABM instances. The resulting mean population sizes were 4587 ($\pm 71$) for prey and 950 ($\pm 40$) for predators, both closely aligned with the target values of $\bar{b}^* = 4575$ and $\bar{c}^* = 948$. (Values in parentheses indicate the unbiased sample standard deviation.) These results show that the controller performs reliably on new data.

Control solutions obtained using the described NNC approach have been compared with those derived from several surrogate models. The results show that surrogate-based control solutions deviate more from the optimum than those produced by the NNC~\cite{fonseca2025optimal}.
\subsection{Inventory dynamics}
\label{sec:inventory}
The following examples focus on neural network\textendash controlled inventory management problems~\cite{gijsbrechts2025ai}, which arise across various industries, including manufacturing, retail, warehousing, and energy. The primary objective in these problems is to determine the optimal ordering policy, which may involve one or more suppliers, that minimizes total costs. These costs usually include ordering expenses, holding costs for excess inventory, and penalties associated with stockouts under stochastic demand.

The sourcing problems we consider are generally formulated as infinite-horizon models aimed at minimizing the expected cost per period under stationary stochastic demand. When training NNCs, we optimize costs across multiple demand trajectories~\cite{bottcher2023control}. This enables us to handle not only non-stationary demand but also both finite-horizon and infinite-horizon discounted settings. Unlike traditional model-free reinforcement learning methods~\cite{gijsbrechts2022can}, the approach pursued here leverages knowledge of the system dynamics, enabling more efficient training and more accurate solutions.

A fundamental yet analytically intractable problem in inventory management is dual sourcing~\cite{barankin1961delivery,fukuda1964optimal,xin2023dual}, where decisions must be made about ordering from either a low-cost, regular supplier or a higher-cost, expedited supplier. In contrast, single sourcing~\cite{arrow1951optimal,scarf1958inventory} is analytically tractable and often serves as a baseline for evaluating policies in more complex multi-sourcing scenarios.

In the following two sections, we apply NNCs to learn ordering policies for single- and dual-sourcing problems. Our implementation is provided in the Python library \texttt{idinn}, which supports both traditional and neural policies for these sourcing models.\footnote{Documentation and examples are available at \url{https://inventory-optimization.readthedocs.io/en/latest/}.}
\subsubsection{Single-sourcing problems}
\begin{figure*}
    \centering
    \includegraphics{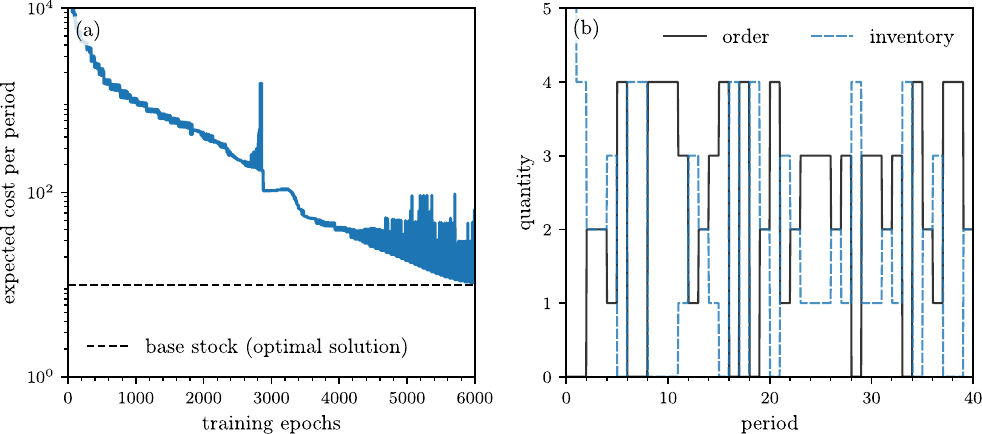}
    \caption{Controlling single-sourcing dynamics using an NNC. (a) Training performance of the learned ordering policy, showing the expected cost per period over training epochs for single-sourcing dynamics with lead time $l = 0$, unit holding cost $h = 5$, unit shortage cost $b = 495$, and demand distribution $\mathcal{U}\{0, 4\}$. The dashed black line indicates the optimal cost of 10 achieved by the base-stock policy. The simulation time was approximately 2 minutes on a standard laptop. (b) Example trajectory of the learned policy, illustrating order quantities (solid black line) and inventory levels (dashed blue line) over time.}
    \label{fig:single_sourcing}
\end{figure*}
In single-sourcing dynamics, the net inventory evolves according to
\begin{equation}
    I_{k+1} = I_{k} + q_{k-l} - D_k \,,
\end{equation}
where $q_{k-l}$ is the replenishment order placed $l$ periods earlier (\ie, the order arriving in period $k\in\{0,1,2,\dots\}$, $l$ is the lead time, and $D_k$ denotes the stochastic demand in period $k$.\footnote{We use the term ``period'' rather than ``time step'' in the context of inventory dynamics, as it aligns more closely with the standard terminology in that field.}

The cost incurred in period $k$ is
\begin{equation}
    c_k = h \max\{0, I_{k+1}\} + b \max\{0, -I_{k+1}\} \,,
\label{eq:single_sourcing_cost}
\end{equation}
where $h$ and $b$ denote the unit holding and shortage costs, respectively. The term $\max\{0, I_{k+1}\}$ represents excess inventory (on-hand stock), while $\max\{0, -I_{k+1}\}$ captures inventory shortages (backorders). The objective is to minimize the total expected cost accumulated over time.

To mathematically describe the optimal ordering policy in single-sourcing problems~\cite{arrow1951optimal,scarf1958inventory}, we let $z$ denote the target inventory position (\ie, the desired net inventory level plus all outstanding orders). The inventory position at time $t$ under single-sourcing dynamics, $\tilde{I}_t$, is
\begin{equation}
    \tilde{I}_k=
    \begin{cases}
    I_k\,,\quad &\text{if} \,\, l=0\\
    I_k+\sum_{i=1}^l q_{t-i}\,, \quad &\text{if} \,\, l>0\,.
    \end{cases}
\end{equation}
The optimal target inventory level~\cite{arrow1951optimal} is given by the critical fractile
\begin{equation}
    z^* = \Phi^{-1}\left(\frac{b}{b+h}\right)\,,
\end{equation}
where $\Phi(x) = \Pr(D \leq x)$ is the cumulative distribution function of demand $D$ over $l+1$ periods. If the inventory position in period $k$ falls below $z^*$, a replenishment order $q_k = z^* - \tilde{I}_k$ is placed to bring the inventory position back to the optimal target level. The optimal single-sourcing policy, often referred to as the ``base-stock policy'', is
\begin{equation}
    q_k=\max\{0,z^*-\tilde{I}_k\}\,,
\label{eq:optimal_base_stock}
\end{equation}
that is, the positive part of $z^* - \tilde{I}_k$. This quantity depends on the optimal inventory position $z^*$, the current net inventory, and the sum of past orders $q_{k-i}$ for $i \in\{1,\dots,l\}$, where $l>0$.

Notice that the mathematical structure of the base-stock policy resembles that of a rectified linear unit (ReLU). The ReLU function returns the positive part of a real number. That is, ${\rm ReLU}(x) = \max\{0, x\}$.

To build an NNC that learns replenishment orders $\hat{q}_k$, we use $l+1$ inputs representing the current net inventory and previous orders (\ie, the system state). We also include a bias term in the input layer to capture the unknown optimal target inventory level $z^*$. These inputs are passed through a ReLU-type activation function that generalizes the expression in Eq.~\eqref{eq:optimal_base_stock}.

However, during training with backpropagation, a ReLU unit can become inactive and consistently output values near zero, often due to a large negative bias term~\cite{DBLP:conf/acssc/DouglasY18}. Once this happens, the corresponding gradients vanish, making it difficult for gradient descent to update the weights. As a result, the output remains stuck near zero, a phenomenon known as the ``dead ReLU'' problem. To avoid this issue, we instead use a continuously differentiable exponential linear unit
\begin{align}
    {\rm CELU}(x;\alpha)=&\max\{0,x\}- \max\{0,\alpha \left(1-\exp(x/\alpha)\right)\}\,,
\label{eq:celu}
\end{align}
which approaches ${\rm ReLU}(x)=\max\{0,x\}$ in the limit $\alpha\rightarrow 0^+$~\cite{barron2017continuously}. CELUs offer an advantage over ReLUs because their smooth, continuous derivatives make gradient calculations more stable during neural network training.

As in the predator-prey example discussed in Section~\ref{sec:predator_prey}, we employ a straight-through estimator to produce integer-valued order quantities while enabling backpropagation of real-valued gradients [see Eq.~\eqref{eq:control_two_node}]. The loss function we optimize is the expected cost per period
\begin{equation}
    J_3[\{c^{(j)}_k\}]=\frac{1}{T}\sum_{k=0}^{T-1} \gamma^k \frac{1}{M}\sum_{j=1}^M c^{(j)}_k\,,
\label{eq:sourcing_loss}
\end{equation}
where $\gamma$ is a discount factor (set to 1 in our case), $M$ is the number of realizations of the sourcing dynamics, and $c^{(j)}_k$ is the cost in period $k$ for the $j$-th realization [see Eq.~\eqref{eq:single_sourcing_cost}].

As an example, we consider a single-sourcing problem with lead time $l = 0$, unit holding cost $h = 5$, unit shortage cost $b = 495$, and discrete uniform demand distribution $ \mathcal{U}\{0, 4\}$. For these parameters, the optimal order-up-to level is $z^* = 4$, and the corresponding optimal expected cost per period is $h(z^* - \bar{D}) = 10$, where $\bar{D} = 2$ is the mean demand per period.
\begin{figure*}
    \centering
    \includegraphics{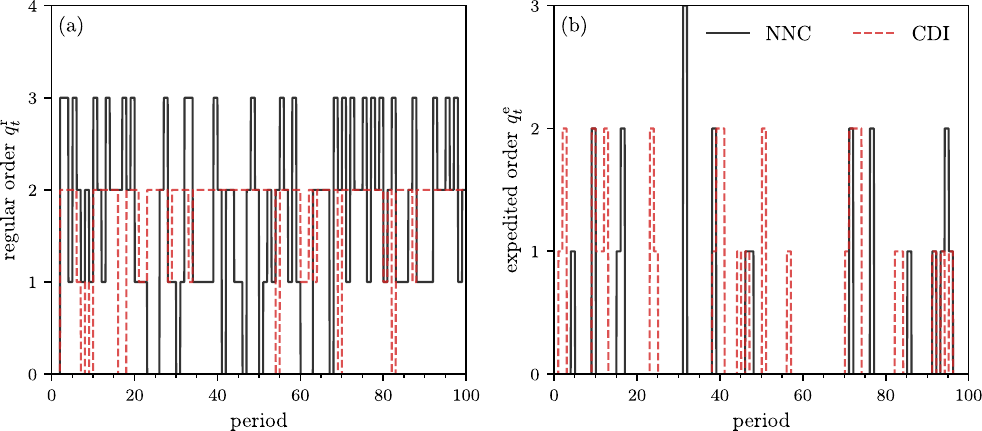}
    \caption{Evolution of regular orders (a) and expedited orders (b) under NNC and CDI policies. The underlying dual-sourcing problem is defined by parameters $h = 5$, $b = 495$, $c_{\rm r} = 0$, $c_{\rm e} = 20$, $l_{\rm r} = 2$, and $l_{\rm e} = 0$. Demand follows a discrete uniform distribution $\mathcal{U}\{0,4\}$.}
    \label{fig:nnc_cdi_order_comparison}
\end{figure*}

For training the NNC, we set $T = 50$ and use $M = 128$ realizations. We observe that the NNC approaches the expected cost level of the optimal base-stock policy after approximately 6000 training epochs [see Fig.~\ref{fig:single_sourcing}(a)]. The total training time is approximately 2 minutes on a standard laptop. As the NNC converges toward the optimal solution, small changes in the neural network weights can lead to large fluctuations in the expected cost, resulting in the onset of oscillatory behavior after around 4500 epochs.

By extracting the NNC parameters corresponding to the lowest observed cost, we can examine the evolution of inventory and order quantities in a representative trajectory [see Fig.~\ref{fig:single_sourcing}(b)]. We find that the NNC successfully learns the optimal base-stock policy, placing orders that consistently reach the optimal order-up-to level $z^* = 4$. This is also reflected in the learned NNC parameters, which align with those of a base-stock policy.
\subsubsection{Dual-sourcing problems}
Building on the previous example of single-sourcing dynamics, we now turn to dual-sourcing problems, which are generally analytically intractable. In such problems, the first sourcing option is a ``regular'' supplier, $\mathrm{R}$, which delivers goods with a (non-negative) integer lead time $l_{\mathrm{r}} > 0$ at a cost $c_{\mathrm{r}}$. A second option is an ``emergency'' supplier, $\mathrm{E}$, which provides goods with a shorter lead time $l_{\mathrm{e}} < l_{\mathrm{r}}$ but at a higher cost $c_{\mathrm{e}} > c_{\mathrm{r}}$. The premium paid for expedited delivery via supplier $\mathrm{E}$ is thus defined as $c \coloneqq c_{\mathrm{e}} - c_{\mathrm{r}} > 0$.

As in the single-sourcing setting, we denote by $I_k$ the net inventory at the beginning of period $k$. The replenishment orders placed in period $k$ to suppliers $\mathrm{E}$ and $\mathrm{R}$ are denoted by $q_k^{\mathrm{e}}$ and $q_k^{\mathrm{r}}$, respectively. In each period $k$, stochastic demand $D_t$ is realized.

Using these definitions, the sequence of events in each period of the dual-sourcing model is as follows:
\begin{enumerate}
    \item At the beginning of period $k$, the inventory manager places replenishment orders $q_k^{\mathrm{r}}$ and $q_k^{\mathrm{e}}$, based on the current net inventory $I_k$ and the outstanding orders that have not yet arrived: $Q_k^{\mathrm{r}} = (q_{k-l_{\mathrm{r}}}^{\mathrm{r}}, \dots, q_{k-1}^{\mathrm{r}})$ and $Q_k^{\mathrm{e}} = (q_{k-l_{\mathrm{e}}}^{\mathrm{e}}, \dots, q_{k-1}^{\mathrm{e}})$.
    \item Orders $q_{k-l_{\mathrm{r}}}^{\mathrm{r}}$ and $q_{k-l_{\mathrm{e}}}^{\mathrm{e}}$ arrive and are added to the inventory.
    \item Demand $D_k$ is realized and subtracted from the current inventory.
\end{enumerate}
The dual-sourcing problem is a Markov decision process with state $(I_{k},Q_{k}^{\rm r},Q_{k}^{\rm e})$ and action $(q_k^{\rm r},q_k^{\rm e})$, where the inventory level (including backlogged excess demand) evolves according to
\begin{equation}
    I_{k+1} = I_k + q_{k-l_{\mathrm{r}}}^{\mathrm{r}} + q_{k-l_{\mathrm{e}}}^{\mathrm{e}} - D_k \,.
    \label{eq:inv_update}
\end{equation}
The corresponding total cost in period $k$ is
\begin{align}
\begin{split}
    c_k=&c_{\rm r}q^{\rm r}_k+c_{\rm e}q^{\rm e}_k+h\max\{0,I_k+q^{\rm r}_{k-l_{\rm r}}+q^{\rm e}_{k-l_{\rm e}}-D_k\} \\
    &+ b\max\{0,D_k-I_k-q^{\rm r}_{k-l_{\rm r}}-q^{\rm e}_{k-l_{\rm e}}\}\,.
\end{split}
\label{eq:dual_sourcing_cost}
\end{align}

A stationary optimal policy exists for minimizing the expected cost per period when the demand distribution has finite support over the considered time horizon~\cite{hua2015structural}. However, a general analytical characterization of the optimal policy in dual-sourcing problems remains elusive.
\begin{figure*}
    \centering
    \includegraphics[width=0.64\textwidth]{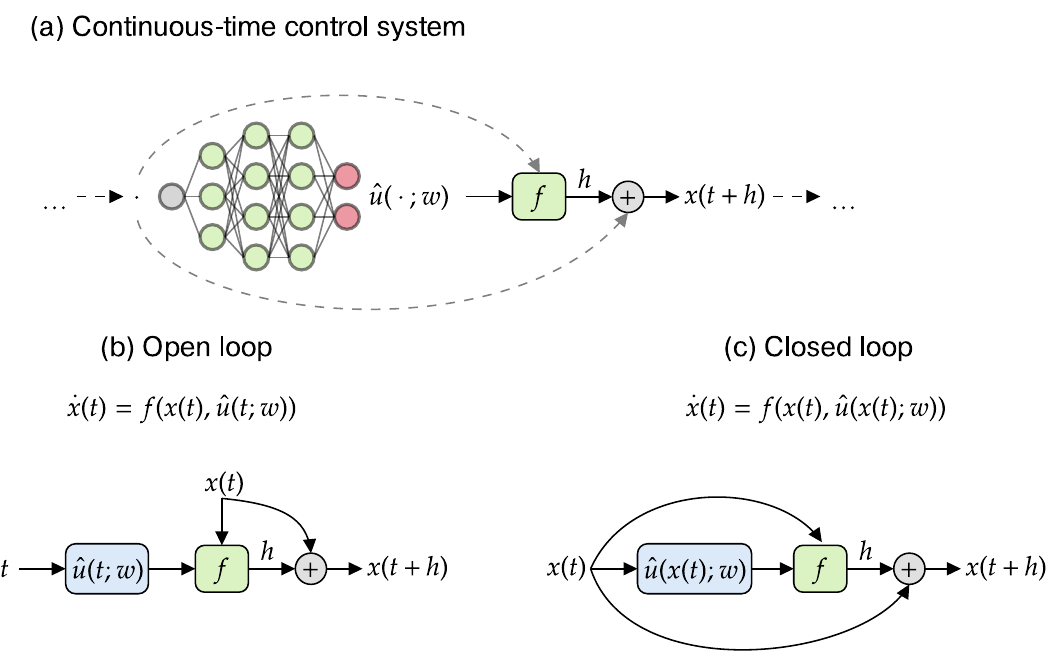}
    \caption{Schematic of a continuous-time control problem, where a control input $\hat{u}(\cdot; w)$ is parameterized by a neural network with parameters $w\in\mathbb{R}^N$. We refer to these control inputs as neural ODE controllers (NODECs). (a) The overall system structure, in which a NODEC generates control inputs that influence the vector field $f(\cdot)$. (b) An open-loop control scenario, where the control input $\hat{u}(t; w)$ depends on the time $t$ but not on the system state $x(t)$. (c) A closed-loop control scenario, where the control input $\hat{u}(x(t); w)$ is computed based on the current state $x(t)$. In both control scenarios, the future state $x(t + h)$ is computed by integrating the system dynamics over the time step $h$. For example, using an explicit Euler scheme, the function $f(\cdot)$ is used to determine the rate of change $\dot{x}(t)$, which is then multiplied by $h$ and added to the current state $x(t)$ to obtain the next state (\ie, $x(t+h)=x(t)+hf(x(t),\hat{u}(\cdot;w))$).}
    \label{fig:continuous_time_control}
\end{figure*}

As an example, we consider a dual-sourcing problem with parameters $h = 5$, $b = 495$, $c_{\rm r} = 0$, $c_{\rm e} = 20$, $l_{\rm r} = 2$, and $l_{\rm e} = 0$. The demand follows a discrete uniform distribution $\mathcal{U}\{0,4\}$. For this instance, the best-performing available heuristic, the capped dual index (CDI) policy~\cite{sun2019robust}, achieves an expected cost per period of 23.26, while the optimal expected cost per period, determined via dynamic programming, is 23.07. 

For the NNC architecture, we use seven hidden layers with 128, 64, 32, 16, 8, 4, and 2 CELU neurons, respectively [see Eq.~\eqref{eq:celu} with $\alpha = 1$]. We train the NNC using the loss function \eqref{eq:sourcing_loss}, based on the dual-sourcing cost \eqref{eq:dual_sourcing_cost}. We consider a time horizon of $T = 1000$ and use $M = 500$. The NNC policy achieves an expected cost per period of 23.13. Extensive testing of NNC policies over a large set of instances has demonstrated that it performs as well as or better than CDI~\cite{bottcher2023control}.

In Fig.~\ref{fig:nnc_cdi_order_comparison}, we show the evolution of regular and expedited orders under the CDI and NNC policies. A key advantage of NNC over CDI is its ability to tailor regular order decisions to the current inventory level and past order placements. As a result, the need for expedited orders is reduced.
\section{Continuous-time dynamics}
\label{sec:cont_time}
We now turn our focus to continuous-time dynamical systems of the form
\begin{equation}
    \dot{x} = f(x, u)\,,
\end{equation}
where $x \equiv x(t) \in \mathbb{R}^n$ and $u \equiv u(t) \in \mathbb{R}^m$ represent the system state and control input at time $t$, respectively. The vector field is $f \colon \mathbb{R}^n \times \mathbb{R}^m \rightarrow \mathbb{R}^n$. We consider both open-loop and closed-loop control inputs, $\hat{u}(t; w)$ and $\hat{u}(x(t); w)$, parameterized by a neural network with parameters $w \in \mathbb{R}^N$. We refer to these controllers as neural ODE controllers (NODECs). As in the discrete-time case, we train these controllers using standard optimizers such as Adam and RMSProp, as implemented in \texttt{PyTorch}.

In Fig.~\ref{fig:continuous_time_control}, we show a schematic of a continuous-time control problem with a NODEC $\hat{u}(\cdot; w)$. In Fig.~\ref{fig:continuous_time_control}(b), we illustrate an open-loop control scenario, where the control input $\hat{u}(t; w)$ depends on time $t$ but not on the system state $x(t)$. In contrast, Fig.~\ref{fig:continuous_time_control}(c) depicts a closed-loop control scenario, where the control input $\hat{u}(x(t); w)$ depends on the current state $x(t)$, allowing the NODEC to adapt its output based on the system’s behavior.

To train NODECs, we define and optimize a suitable loss function that reflects the control objective. Analogous to the discrete-time finite-horizon cost functional in Eq.~\eqref{eq:discrete_time_cost}, a commonly used continuous-time counterpart is
\begin{equation}
    J[x(t), u(t)] = \int_0^T L(x(t), u(t))\,\mathrm{d}t + V(x(t))\,,
    \label{eq:continuous_time_cost}
\end{equation}
where $L(x(t), u(t))$ denotes the running cost at time $t$, and $V(x(t))$ is the terminal cost associated with the final state. By replacing $u(\cdot)$ with the parameterized control input $\hat{u}(\cdot; w)$, we optimize the cost functional $J[x(t), \hat{u}(\cdot)]$ using automatic differentiation to update the NODEC parameters $w$.

We focus on several examples to demonstrate the applicability of NODEC to continuous-time dynamical systems. In the first example, we examine a basic control problem involving the movement of a block subject to friction, where the objective is to minimize work~\cite{gong2006pseudospectral,bottcher2022near}. Next, we apply NODEC to Lotka--Volterra-type predator-prey dynamics, which arise, for instance, in biomedical contexts such as modeling microbiome interactions~\cite{faust2012microbial}. Finally, we consider the control of systems composed of coupled oscillators~\cite{bottcher2022ai}.
\subsection{Illustrative example}
\begin{figure}
    \centering
    \includegraphics[width=\linewidth]{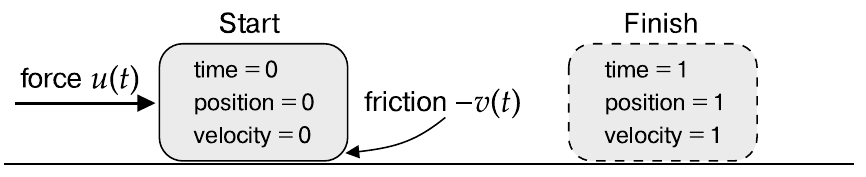}
    \caption{Schematic of the block-move example.}
    \label{fig:block_move}
\end{figure}
As an illustrative continuous-time control problem, we consider the task of moving a block in the presence of friction (see Fig.~\ref{fig:block_move}). The objective is to minimize the total work
\begin{equation}
    W[u,v]=\int_0^T u(t)v(t) \,\mathrm{d}t\,,
\label{eq:work}
\end{equation}
subject to the dynamics and constraints
\begin{align}
\begin{split}
    \dot{x}(t)      &= v(t)\,, \\
    \dot{v}(t)      &= -v(t) + u(t)\,, \\
    v(t)            &\geq 0\,, \\
    0               \leq u(t) &\leq 2\,, \\
    (x(0), v(0))    &= (0, 1)\,, \\
    (x(T), v(T))    &= (1, 1)\,,
\end{split}
\end{align}
where the time horizon is $T=1$~\cite{gong2006pseudospectral}. The control input $u(t)$ represents the force applied to the block, and in this case, the optimal control is $u^*(t) = 1$. Although the solution is analytically straightforward, some numerical methods have difficulty accurately approximating the constant control input~\cite{gong2006pseudospectral}.
\begin{figure*}[t]
    \centering
    \includegraphics{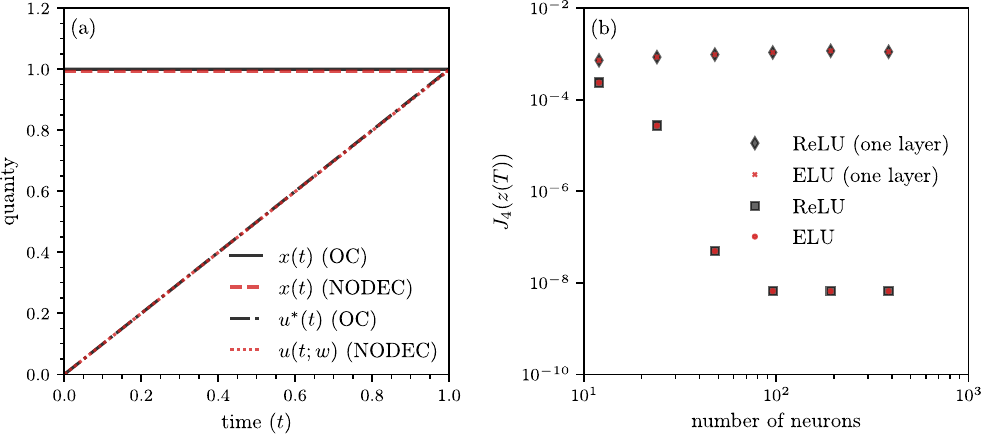}
    \caption{Control of a moving block. (a) Evolution of the state $x(t)$ and control input $u(t)$. Optimal control (OC) and NODEC-based solutions are shown in black and red, respectively. (b) The loss $J_4(z(T)) = \|z(T) - z^*\|_2^2$ for different activation functions and architectures, plotted as a function of the number of neurons. Diamonds and crosses indicate solutions based on single-layer architectures. For square and disk markers, the number of hidden layers is 2, 4, 8, 16, or 32, with 6 neurons per layer.}
    \label{fig:moving_particle}
\end{figure*}

To solve this control problem using NODEC, we represent $u(t)$ with a neural network $\hat{u}(t; w)$ consisting of eight hidden layers, each containing six ELU neurons. The controller is trained for 100 epochs using the Adam optimizer with a learning rate of $5\times 10^{-3}$, minimizing the loss
\begin{equation}
    J_4(z(T)) = \|z(T) - z^*\|_2^2\,,
    \label{eq:loss_J4}
\end{equation}
where $z(T) = (x(T), v(T))^\top$ denotes the final system state, and $z^* = (1, 1)^\top$ is the desired target state. That is, we do not explicitly minimize the work $W[u,v]$.

In Fig.~\ref{fig:moving_particle}(a), we show the evolution of $x(t)$ and $u(t)$. The optimal control (OC) and the NODEC solutions are shown as grey and red lines, respectively. We observe that NODEC has learned a control input that closely resembles the optimal one.

We now study how the structure of the neural network and the choice of activation function affect the performance of NODEC. To this end, we fix the number of neurons per layer to six and vary the number of hidden layers from $2$ to $64$, initializing all weights and biases to $10^{-2}$. Controllers are trained for $100$ epochs using the Adam optimizer with a learning rate of $5 \times 10^{-3}$, and the best-performing model is selected based on the loss value. In Fig.~\ref{fig:moving_particle}(b), we compare results for both ELU and ReLU activations in terms of the final loss value. For networks with at least $8$ hidden layers, the loss consistently drops below $10^{-7}$. In contrast, single-layer networks, regardless of width, fail to approximate the optimal control solution. Overall, performance is nearly identical for both ELU and ReLU activation functions.

In addition to the previous numerical experiment, we now explicitly include the additional loss term $W[u,v]$ and optimize the objective
\begin{equation}
    J_5[u,v]=\|z(T) - z^*\|_2^2+\mu W[u,v]\,,
\label{eq:loss_2}
\end{equation}
where $\mu$ is a Lagrange multiplier that determines the influence of $W[\cdot]$ on the total loss. To initialize the weights, we use the Kaiming uniform initializer~\cite{he2015delving}, and we set all biases initially to a value of $10^{-2}$.
\begin{figure*}
    \centering
    \includegraphics{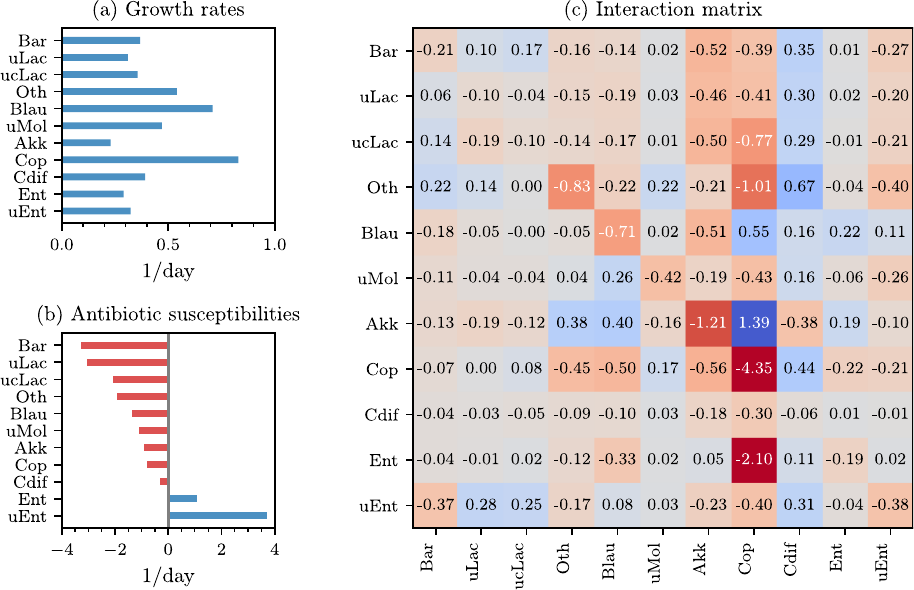}
    \caption{Microbial dynamics under antibiotic perturbation. (a) Species-specific growth rates (1/day). (b) Antibiotic susceptibilities (1/day) in response to clindamycin, with negative values (red) indicating inhibition and positive values (blue) indicating facilitation. (c) Interaction matrix quantifying pairwise effects between species (element $m_{ij}$ shows effect of species $j$ on species $i$), with negative values (red) indicating inhibition and positive values (blue) indicating facilitation. Species are ordered by antibiotic susceptibility from most inhibited to most promoted. Species abbreviations: Bar (\textit{Barnesiella}), uLac (undefined Lachnospiraceae), ucLac (unclassified Lachnospiraceae), Oth (Other), Blau (\textit{Blautia}), uMol (unclassified Mollicutes), Akk (\textit{Akkermansia}), Cop (\textit{Coprobacillus}), Cdif (\textit{C.~difficile}), Ent (\textit{Enterococcus}), uEnt (undefined Enterobacteriaceae). Data is based on \cite{buffie2012profound,stein2013ecological}.}
    \label{fig:glv_data}
\end{figure*}

The neural network that we use to optimize $J_5[u,v]$ consists of $8$ hidden layers, each containing $6$ ELU neurons. We train the network using the Adam optimizer with a learning rate of $\eta = 10^{-1}$ for 100 epochs, and we evaluate the best-performing model.

The use of uniform weight initialization and a relatively large learning rate represents a deliberately non-optimized hyperparameter configuration that requires tuning of the multiplier $\mu$.

As shown in~\cite{bottcher2022near}, the loss term $\|z(T) - z^*\|_2^2$ reaches a minimum for $\mu \approx 2 \times 10^{-3}$, and the corresponding value of $W[u,v]$ is reasonably close to the optimal solution. However, the results in~\cite{bottcher2022near} also demonstrate that implicit regularization by excluding $W[u,v]$ from the loss [see Eq.~\eqref{eq:loss_J4}] can achieve lower overall loss values while still producing similar values of $W[u,v]$.

Furthermore, when analyzing solutions obtained for different values of the multiplier $\mu$, we observe that while the state trajectories $x(t)$ remain largely aligned with the optimal control solution, variations in $\mu$ lead to noticeable deviations in the control input $\hat{u}(t; w)$ relative to the optimal control $u^*(t)$.
\subsection{Population dynamics}
\label{sec:predator_prey_cont}
We continue our discussion of NODEC by applying it to models of population dynamics, which are useful for studying species interactions in both ecological and biomedical contexts. While we examined discrete-time formulations for single- and multi-species dynamics in Sections~\ref{sec:discrete_time_example} and~\ref{sec:predator_prey}, we now shift our focus to a continuous-time framework and consider the generalized Lotka--Volterra (gLV) equations. This model has been employed, for instance, in microbial ecology~\cite{faust2012microbial}, where species frequently engage in inhibitory and facilitative interactions. In this context, a commonly used form of the gLV equations is
\begin{equation}
\dot{x}_i(t) = x_i(t) \left( b_i + \sum_{j=1}^{n} m_{ij} x_j(t)+\epsilon_i u(t)\right)\,,
\end{equation}
where $x_i(t)$ denotes the abundance of microbial species $i$ at time $t$, $b_i$ is its intrinsic growth rate, and $m_{ij}$ represents the interaction coefficient quantifying the effect of species $j$ on species $i$. The term $\epsilon_i u(t)$ models the effect of an external antibiotic treatment $u(t)$, where $\epsilon_i$ indicates the antibiotic susceptibility of species $i$. A negative $\epsilon_i$ corresponds to inhibition by the antibiotic, while a positive value indicates the opposite.
\begin{figure*}
    \centering
    \includegraphics{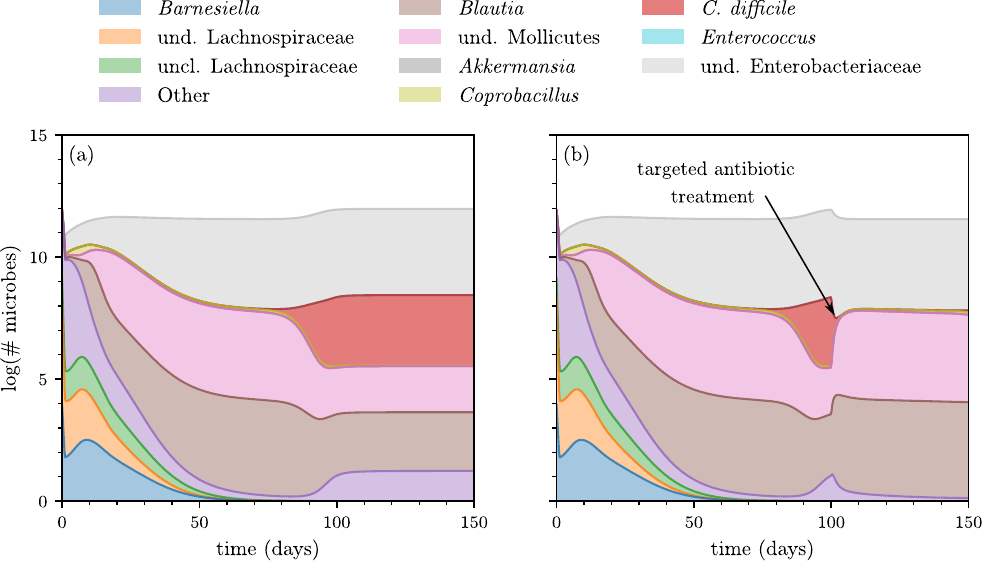}
    \caption{Simulated microbial dynamics under antibiotic interventions. (a) Without targeted treatment, the initial administration of clindamycin promotes the outgrowth of \textit{C.~difficile}, leading to a persistent infection. (b) NODEC-based targeted antibiotic treatment starting on day 100 (black arrow), effectively suppressing \textit{C.~difficile}. Colors represent different microbial groups as indicated in the legend.}
    \label{fig:antibiotic_simulation}
\end{figure*}

While gLV models are widely used to study microbial systems, a key limitation is their assumption of direct interactions between microbial species, which overlooks indirect effects mediated by competition for shared nutrients.

In \cite{stein2013ecological}, gLV parameters were inferred using mouse data from a study~\cite{buffie2012profound} that examined the effect of the antibiotic clindamycin on intestinal colonization by the spore-forming pathogen \textit{C.~difficile}. The dataset includes a total of $n=11$ species. In Fig.~\ref{fig:glv_data}, we show the estimated growth rates $b_i$, clindamycin susceptibilities $\epsilon_i$, and elements of the interaction matrix $m_{ij}$. All growth rates are positive, while the diagonal elements of the interaction matrix, $m_{ii}$, are negative. These negative values indicate that each species can reach its carrying capacity even in the absence of other species. The inferred clindamycin susceptibilities suggest that the antibiotic inhibits all microbial species except \textit{Enterococcus} and an undefined group of \textit{Enterobacteriaceae}. \textit{C.~difficile} itself appears to be only mildly inhibited by clindamycin.

We now consider a control problem focused on treating a \textit{C.~difficile} infection following the administration of clindamycin~\cite{jones2018silico,jones2020navigation,bonnard2024geometric}.

We use the growth rates, interaction coefficients, and clindamycin susceptibilities from \cite{buffie2012profound,stein2013ecological} (see Fig.~\ref{fig:glv_data}) and initialize the model with initial condition 5 from \cite{stein2013ecological,jones2018silico}. To simulate infection onset, we introduce a small initial perturbation of $10^{-10}$ (in nondimensional units) to the \textit{C.~difficile} compartment and apply a unit dose of clindamycin on the first day. This treatment protocol is consistent with the constant dosing schedule considered in \cite{jones2018silico}.

In Fig.~\ref{fig:antibiotic_simulation}(a), we show the corresponding evolution of microbial species. The results indicate that the initial antibiotic intervention, in combination with the \textit{C.~difficile} perturbation, leads to a substantial infection after approximately 90--100 days. In the absence of both perturbation and treatment, the system evolves as in Fig.~3(a) of \cite{jones2018silico}.

\textit{C.~difficile} infections are, for instance, treated with antibiotics such as vancomycin or metronidazole~\cite{mcfarland2002breaking}. Following the approach in~\cite{jones2018silico}, we now consider a hypothetical targeted antibiotic that is highly effective against \textit{C.~difficile}. The treatment begins on day 100 and lasts for 10 days. In our model, we set the antibiotic susceptibility of \textit{C.~difficile} to $-1$, while the susceptibilities of all other microbial species are set to $0$. We train a NODEC to minimize the loss function
\begin{align}
\begin{split}
    J_6[x,\hat{u}] =& \frac{1}{50} \int_{100}^{150} x_9(t)\,\mathrm{d}t \\
    &+ \mu \frac{1}{10} \int_{100}^{110} \hat{u}^2(t; w)\,\mathrm{d}t\,,
\end{split}
\end{align}
which penalizes the abundance of \textit{C.~difficile} (modeled by compartment $x_9$) over time, while also promoting prudent use of the targeted antibiotic. Time is measured in days. In our simulations, we set $\mu=0.01$.

In Fig.~\ref{fig:antibiotic_simulation}(b), we show the simulation results obtained using the trained controller. We observe that the targeted antibiotic treatment successfully suppresses the \textit{C.~difficile} infection. The NODEC that we employ consists of five hidden layers, each with four ELU neurons. Training was performed for 200 epochs using the Adam optimizer with a learning rate of $10^{-3}$, yielding a minimum loss of 0.061. As a baseline for comparison, we simulated a naive treatment strategy that administered a constant unit dose per day over the same 10-day period. This approach resulted in a loss more than ten times higher and failed to eliminate the infection.
\subsection{Oscillator dynamics}
We conclude this section on continuous-time dynamics by examining control problems associated with the Kuramoto model~\cite{kuramoto1975self}, which describes a system of coupled oscillators. Each oscillator is characterized by a phase $\theta_i$ and an intrinsic (natural) frequency $\omega_i$, where $i \in \{1, \dots, n\}$. The system dynamics are given by
\begin{align}
\begin{split}
    \dot{\Theta}(t) &= \Omega + f(\Theta(t),u(t))\\
    \Theta(0)&=\Theta_0\,,
\label{eq:kuramoto}
\end{split}
\end{align}
where $\Theta = (\theta_1, \dots, \theta_n)^\top$ and $\Omega = (\omega_1, \dots, \omega_n)^\top$~\cite{kuramoto1975self}. We sample the natural frequencies $\omega_i$ and initial phases $\theta_i(0)$ from a normal distribution with mean $0$ and standard deviation $0.2$.

The interactions among oscillators, as well as the effect of control inputs $u_i(t)$ on oscillator $i$, are modeled by the function
\begin{equation}
    f_i(\Theta(t),u(t))=\frac{K u_i(t)}{n} \sum_{j=1}^n a_{ij} \sin(\theta_j(t)-\theta_i(t))\,,
\label{eq:kuramoto2}
\end{equation}
where $K$ is the coupling strength, and $a_{ij}$ are the adjacency-matrix elements representing the underlying undirected network. To evaluate the level of synchronization at the final time $T$, we use the complete synchronization condition
\begin{equation}
    |\dot{\theta}_i(T)-\dot{\theta}_j(T)|=0~\text{for}~(i,j)\in E\,,
\label{eq:complete_synchronization}
\end{equation}
where $E$ denotes the set of edges in the network~\cite{ha2016emergence, biccari2020stochastic}. When condition \eqref{eq:complete_synchronization} is met, all connected oscillators exhibit constant phase differences. 

In the special case where all control inputs are set to 1 (\ie, $u_i(t) = 1$ $\forall i$) the system described by Eq.~\eqref{eq:kuramoto} possesses a unique and stable synchronized state, provided that the coupling strength $K$ exceeds the threshold
\begin{equation}
    K^*=\norm{L^\dagger \Omega}_{E,\infty}\,,
\end{equation}
where $L^\dagger$ denotes the Moore--Penrose pseudo-inverse of the combinatorial graph Laplacian and $\norm{x}_{E,\infty}=\max_{(i,j)\in E}|x_i-x_j|$ is the maximum distance between elements in $x=(x_1,\dots,x_n)^\top$ that are connected via an edge in $E$~\cite{dorfler2013synchronization}. In our simulations, we use a subcritical coupling strength $K = 0.1 K^*$, such that synchronization requires that some $u_i(t)$ must exceed 1 to achieve it.

For a global control input $u(t)$ (\ie, $u_i(t) = u(t)$ $\forall i$), there exists an optimal control $u^*(t)$ that minimizes the cost functional
\begin{align}
    J_7[\Theta(T), u] &= \frac{1}{2} \sum_{i,j} a_{ij} \sin^2(\theta_j(T) - \theta_i(T)) + \frac{\mu}{2} E[u]\,,
    \label{eq:loss_kuramoto}
\end{align}
where the parameter $\mu$ determines the relative weight of the energy regularization term $E[u]=\int_0^T \|u(t)\|_2^2\,\mathrm{d}t$ in the cost function. Minimizing $J_7[\Theta(T), u]$ aligns with the synchronization objective defined in Eq.~\eqref{eq:complete_synchronization}~\cite{biccari2020stochastic}.

The optimal control for this problem can be computed using the adjoint-gradient method (AGM), which combines Pontryagin's maximum principle with gradient descent on $u$~\cite{biccari2020stochastic}. Specifically, the control is updated according to
\begin{equation}
    u^{(\ell+1)}=u^{(\ell)}-\tilde{\eta} \left[\mu u^{(\ell)} + \frac{K}{n} \sum_{i=1}^n \lambda_i \sum_{j=1}^n a_{ij} \sin(\theta_j-\theta_i) \right]\,,
\label{eq:optimal_kuramoto}
\end{equation}
where $\tilde{\eta}$ denotes the AGM learning rate, and the quantity $\lambda = (\lambda_1, \dots, \lambda_n)^\top$ is the solution to the adjoint system
\begin{align}
\begin{split}
    -\dot{\lambda}_i =& -\frac{K u \lambda_i}{n} \sum_{i\neq j} a_{ij} \cos(\theta_j-\theta_i)\\
    &+\frac{K u}{n} \sum_{i\neq j} a_{ij}\lambda_j\cos(\theta_j-\theta_i)\,,
\end{split}
\label{eq:AGM}
\end{align}
with terminal condition $\lambda_i(T)=1/2\sum_{i \neq j} a_{ij} \sin(2\theta_i(T)-2\theta_j(T))$.

We compare the control performance of NODEC applied to Eq.~\eqref{eq:kuramoto} with that of the AGM. The neural controller we employ learns $\hat{u}(t;w)$ based on the loss function \eqref{eq:loss_kuramoto} with a gradient descent in $w$ and without energy regularization term $\mu E[u]/2$. We denote this loss function by
\begin{equation}
    J_8(\Theta(T))=\frac{1}{2}\sum_{i,j} a_{ij} \sin^2(\theta_j(T)-\theta_i(T))\,.
\label{eq:loss_kuramoto_1}
\end{equation}
Since both NODEC and the AGM rely on different optimization procedures with distinct learning rates, we chose learning rates for which
the corresponding order parameter values are approximately equal. As shown in~\cite{bottcher2022ai}, a high degree of synchronization can be achieved by controlling only a fraction of the nodes. That work also demonstrates how a maximum matching approach~\cite{liu2011controllability} can be used to identify driver nodes for controlling linear dynamics involving over $1000$ nodes.

\begin{figure*}
    \centering
    \includegraphics[width=\textwidth]{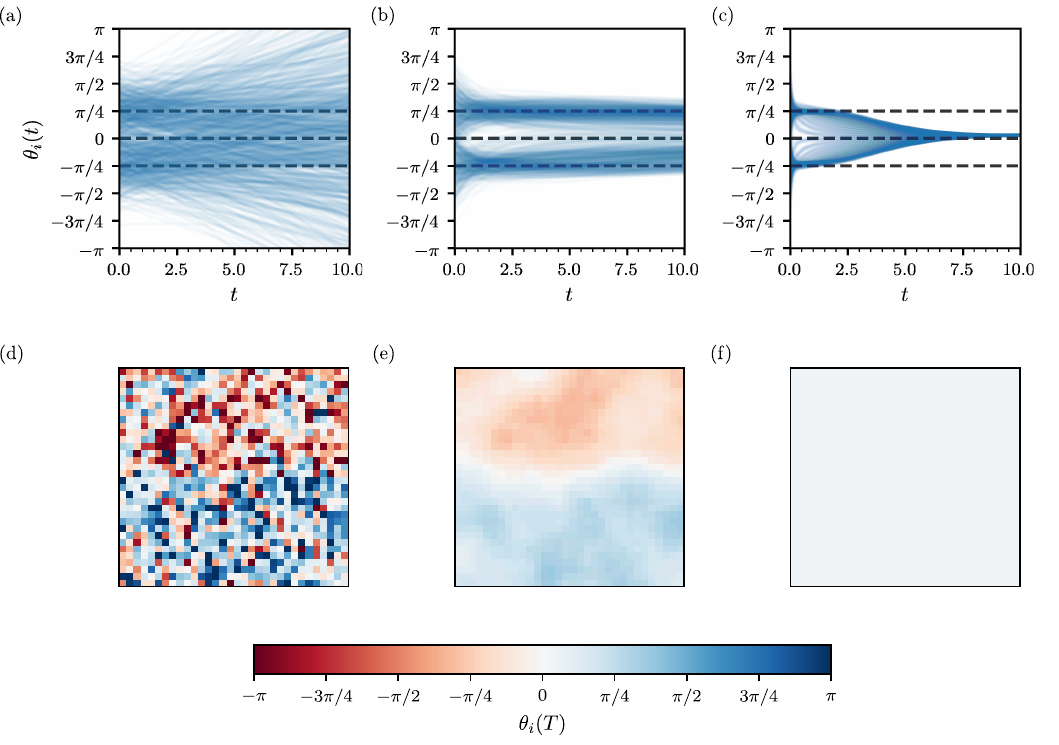}
    \caption{Kuramoto dynamics and distinct target states. We simulate the phase evolution $\theta_i(t)$ of $n = 1024$ coupled Kuramoto oscillators ($i \in \{1,\dots,n\}$), arranged on a $32 \times 32$ square lattice without periodic boundary conditions [see Eqs.~\eqref{eq:kuramoto} and \eqref{eq:kuramoto2}]. A subcritical coupling strength of $K = 0.01 K^*$ is used, and the control horizon is set to $T = 10$. Initially, the oscillator phases follow a bimodal Gaussian distribution with means $-\pi/4$ (top half of the lattice) and $\pi/4$ (bottom half), and variance 0.5. The top panels show the evolution of $\theta_i(t)$, while the bottom panels display the spatial phase distribution $\theta_i(T)$ at the final time $T = 10$. Each pixel in the bottom panels corresponds to the phase of a specific oscillator in the lattice. (a,d) With the control input fixed at $u_i(t) = 1$ for all $i$ (uncontrolled dynamics), the phase differences grow over time.
    (b,e) NODEC successfully drives the system toward a state in which the oscillator phases approach $-\pi/4$ and $\pi/4$ by minimizing the loss function $J_9(\Theta(T))$ [see Eq.~\eqref{eq:loss_kuramoto_2}].
    (c,f) NODEC achieves global synchronization by minimizing $J_8(\Theta(T))$ [see Eq.~\eqref{eq:loss_kuramoto_1}]. The dashed black lines in panels (a--c) indicate reference phases of $-\pi/4$, $0$, and $\pi/4$. The learning rate is set to 15 in panels (b,e) and 0.12 in panels (c,f).}
    \label{fig:target_state_2}
\end{figure*}

A commonly used measure of the degree of synchronization is the order parameter
\begin{equation}
r(t) = \frac{1}{n} \sqrt{\sum_{i,j} \cos \left[\theta_j(t) - \theta_i(t)\right]}.
\label{eq:order_parameter}
\end{equation}
This expression follows from the fact that the squared magnitude of the complex order parameter $z = r e^{i\psi(t)} = \frac{1}{n} \sum_{j=1}^n e^{i\theta_j(t)}$~\cite{kuramoto1975self} can be rewritten as
\begin{equation}
r(t)^2 = |z|^2 = \frac{1}{n^2} \sum_{i,j} \cos\left[\theta_j(t) - \theta_i(t)\right].
\end{equation}
A value of $r(t) = 1$ indicates perfect synchronization, where all oscillators share the same phase.

We now apply NODEC to control oscillator systems on a square lattice with periodic boundaries and with $n = 2500$ nodes, and compare it with the AGM, setting $T = 0.5$. We find that the control energy and order parameter ratios are $E^{\rm NODEC}[u] / E^{\rm AGM}[u] \approx 1.0045$ and $r^{\rm NODEC}(T) / r^{\rm AGM}(T) \approx 0.9999$, respectively. NODEC and the AGM achieve similar values for both the order parameter and control energy at time $T = 0.5$, indicating that both methods effectively control the considered oscillator system. In \cite{bottcher2022ai}, NODEC has also been applied to directed networks, and its robustness to noise has been analyzed.

For a runtime performance comparison, we measure the learning (or wall-clock) time associated with controlling the system. To this end, we determine the runtime of 50 control realizations for both the AGM and NODEC. The mean runtimes are $74~\mathrm{s}$ and $1.03~\mathrm{s}$ for the AGM and NODEC, respectively. For the considered oscillator system, NODEC’s training time is thus approximately two orders of magnitude shorter than that of the AGM. In \cite{bottcher2022ai}, runtime differences between NODEC and the AGM have been analyzed in more detail. The main bottleneck that has been identified in the AGM is that the adjoint system solver requires small step sizes to accurately capture the interaction between the adjoint dynamics [see Eq.~\eqref{eq:AGM}] and the gradient descent updates [see Eq.~\eqref{eq:optimal_kuramoto}] applied to the control inputs.

We now consider a control problem with a target state that differs from full synchronization. Specifically, we aim to steer each oscillator toward either $-\pi/4$ or $\pi/4$. This control objective is captured by the loss function
\begin{equation}
J_9(\Theta(T)) = \frac{1}{2} \sum_{i=1}^n \left( \left| \theta_i(T) \right| - \frac{\pi}{4} \right)^2.
\label{eq:loss_kuramoto_2}
\end{equation}

Figure~\ref{fig:target_state_2} shows that NODEC, when trained using the loss function $J_9(\Theta(T))$, can successfully steer a system of $n=1024$ coupled Kuramoto oscillators, arranged on a square lattice with subcritical coupling strength $K = 0.01 K^*$, towards a target state consisting of two distinct spatial regions. In this configuration, the oscillators converge to phase values $\theta_i(T)$ close to either $-\pi/4$ or $\pi/4$.

In Figs.~\ref{fig:target_state_2}(a,d), no control is applied, and we observe increasing phase dispersion over time. In contrast, Figs.~\ref{fig:target_state_2}(b,e) show that NODEC, trained with $J_9(\Theta(T))$, drives the system toward a target state in which oscillators in the upper half of the lattice converge to phase values near $-\pi/4$ (indicated by light orange), while those in the lower half converge to values around $\pi/4$ (indicated by light blue).

For comparison, we also use NODEC with the synchronization loss function $J_8(\Theta(T))$, which results in complete phase alignment across the lattice as illustrated in Figs.~\ref{fig:target_state_2}(c,f).

In the studied examples, NODEC has two key advantages over adjoint-based control methods. First, approximate optimal control trajectories can be obtained without deriving and solving the adjoint system (see Section~\ref{sec:discrete_time_example} for a corresponding example in discrete time). The only inputs necessary are (i) the dynamical system, (ii) its initial state, and (iii) the desired target state. Second, the runtime of NODEC may be substantially faster than that of adjoint-gradient methods.
\section{Comparison with model predictive control}
\label{sec:comparison_mpc}
\begin{figure*}
    \centering
    \includegraphics{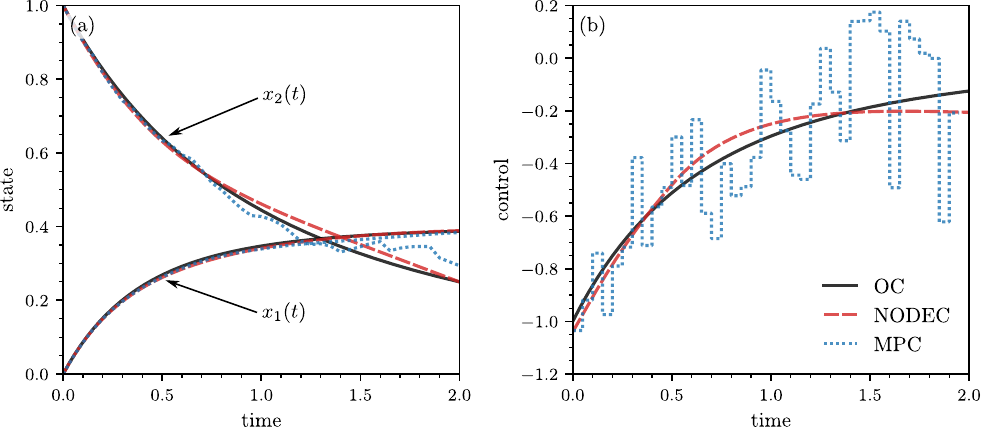}
    \caption{Comparison of control approaches applied to the nonlinear control problem from Eqs.~\eqref{eq:mpc_loss} and \eqref{eq:mpc_1}. (a) Evolution of the state variables $x_1(t)$ and $x_2(t)$ under OC (solid black line), NODEC (dashed red line), and MPC (dotted blue line). (b) Corresponding control inputs. NODEC approximates OC more closely than MPC, which exhibits higher variability.}
    \label{fig:nodec_vs_mpc}
\end{figure*}
Model predictive control (MPC) represents a control paradigm closely related to the neural controllers discussed in previous sections. In MPC, one optimizes a loss function over a finite time horizon at each time step, applies only the first control input, and repeats this process until the final time. This requires solving an optimization problem for each updated system state, which can be computationally demanding, especially in nonlinear settings. However, MPC has the advantage of naturally handling constraints, such as bounds on control inputs, and of adapting in real time to disturbances or model errors. In contrast, NODEC, as employed in the earlier sections, does not solve an optimization problem during execution. Instead, it learns a parameterized control policy by constructing a computational graph over the entire system evolution and minimizing a loss function offline. This approach can be computationally efficient, but enforcing bounds on control inputs or adapting to unexpected conditions would require extensions beyond the approach considered in previous sections.

To provide insights into the different optimization paradigms underlying NODEC and MPC, we adapt a nonlinear optimal control problem from~\cite{gong2006pseudospectral} as our test case. The goal is to minimize
\begin{equation}
J_{10}[u] = 4 \int_0^2 u^2(t) ,\mathrm{d}t\,,
\label{eq:mpc_loss}
\end{equation}
subject to the nonlinear dynamics and boundary conditions
\begin{align}
\begin{split}
\dot{x}_1(t) &= x_2^3(t)\,,\\
\dot{x}_2(t) &= u(t)\,,\\
(x_1(0), x_2(0)) &= (0, 1)\,, \\
(x_1(2), x_2(2)) &= (0.3875, 0.25)\,.
\end{split}
\label{eq:mpc_1}
\end{align}

This control problem admits the analytical solution
\begin{align}
    u^*(t) &= -\frac{8}{(2 + t)^3}\,,\label{eq:mpc_2_1} \\
    x^*_1(t) &= \frac{2}{5} - \frac{64}{5(2 + t)^5}\,,\label{eq:mpc_2_2}\\
    x^*_2(t) &= \frac{4}{(2 + t)^2}\,.\label{eq:mpc_2_3}
\end{align}

To solve the described control problem with NODEC, we use a parameterized control input $\hat{u}(t; w)$, represented by a neural network with two hidden layers, each containing four ELU activations. We trained NODEC using the Adam optimizer with a learning rate of 0.1.

Training was performed in two stages. In the first stage, we minimized the mean squared error (MSE) between the predicted and target values of $x_1(t)$. In the second stage, we jointly minimized the MSE with respect to both $x_1(t)$ and $x_2(t)$. Directly optimizing the full loss, including the control objective $ J_{10}$, did not lead to satisfactory solutions. This observation is consistent with related findings in~\cite{bottcher2022ai}.

For comparison with NODEC, we consider a receding-horizon MPC approach. At each time step $t_k=k\Delta t$ of the discretized dynamics~\eqref{eq:mpc_1}, we solve the finite-horizon optimal control problem
\begin{align}
\begin{split}
    \min_{\{u_k\}} \quad & 100 \left\| z_{N_{\mathrm{p}}} - z^* \right\|^2 + 10 \sum_{k=1}^{N_{\mathrm{p}}-1} (u_k - u_{k-1})^2 \\
    & + \Delta t \sum_{k=0}^{N_{\mathrm{p}}-1} u_k^2\,,
\end{split}
\label{eq:loss_mpc}
\end{align}
where $N_{\mathrm{p}}$ is the number of steps in the prediction horizon, $z^* = (0.3875, 0.25)^\top$ is the desired terminal state, and $z_{N_{\mathrm{p}}}$ denotes the predicted state at the end of the horizon. The multipliers in Eq.~\eqref{eq:loss_mpc} have been chosen to obtain a solution that is aligned with the optimal one.

The state trajectory evolves according to the discretized dynamics
\begin{equation}
    z_{k+1} = f_{\mathrm{d}}(z_k, u_k), \quad z_0 = z(t_k),
\end{equation}
where $f_{\mathrm{d}} \colon \mathbb{R}^n \times \mathbb{R}^m \rightarrow \mathbb{R}^n$ is the discrete-time approximation of the continuous-time system~\eqref{eq:mpc_1} obtained via numerical integration.

After solving the optimization problem, only the first control input $u_0$ is applied to the system. At the next time step, the procedure is repeated using the updated system state. For each time step, we use the Adam optimizer with a learning rate of 0.5 to optimize the control sequence over the finite prediction horizon.

In Fig.~\ref{fig:nodec_vs_mpc}, we show a comparison of the evolution of both the system state and control inputs obtained using OC [see Eqs.~\eqref{eq:mpc_2_1}--\eqref{eq:mpc_2_3}], NODEC, and MPC. NODEC closely tracks the optimal trajectory, whereas MPC exhibits larger deviations and higher variability in the control signal due to its locally optimal receding-horizon updates. NODEC reaches a state of $(0.39, 0.25)$, which is closer to the target state of approximately $(0.39, 0.25)$ than MPC's $(0.38, 0.29)$. Furthermore, NODEC achieves a loss value of 1.56, nearly matching the optimal 1.55, while MPC yields a higher loss of 1.73.

The example presented here is intended to illustrate the different optimization procedures underlying NODEC and MPC. (A related comparison between MPC and a neural control approach using experimental data of CartPole and F1TENTH Race Car systems has been studied in~\cite{paluch2024hardware}.) Despite these differences between neural controllers and MPC, there is significant potential in integrating neural approximators into MPC frameworks. For example, neural ODEs have been employed to model unknown dynamical systems~\cite{saint1991neural,draeger1995model}, or to augment physics-based models in real-world systems such as aerial robots~\cite{chee2022knode}. Additionally, input convex neural networks~\cite{DBLP:conf/icml/AmosXK17} have been used to model systems while preserving convexity properties, enabling efficient optimization in MPC~\cite{DBLP:conf/l4dc/BunningSABHL21}. A recent application of such an approach focuses on brain stimulation in Parkinson's disease~\cite{steffen2025deep}.

Beyond augmenting models with neural approximators, other works have studied the integration of learned dynamics into MPC approaches more broadly. In low-data regimes where interpretability and online adaptability are important, the sparse identification of nonlinear dynamics (SINDY) framework~\cite{kaiser2018sparse} provides a promising alternative to standard neural approximators in MPC tasks. In another work, researchers have proposed a two-stage framework that first learns the dynamics of networked systems offline using graph neural networks, and then employs MPC online using the learned model to compute control inputs~\cite{mou2024model}. This approach has been applied to complex systems including an agent-based model, a networked epidemic model, and the Kuramoto model.
\section{Uncertainty quantification with conformal prediction}
\label{sec:conformal_prediction}
\begin{figure}
    \centering
    \includegraphics{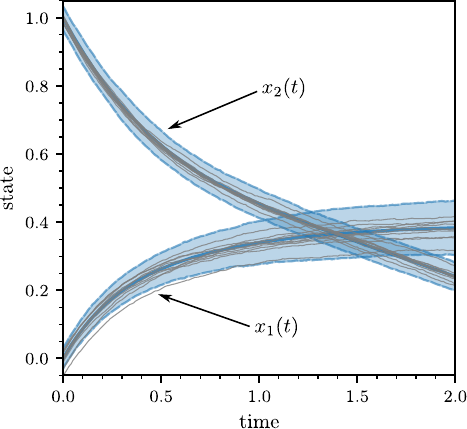}
    \caption{Conformal-prediction intervals for NODEC-controlled trajectories of the nonlinear system in Eqs.~\eqref{eq:mpc_loss}--\eqref{eq:mpc_1}, under process and initial state noise. The solid blue line indicates the mean trajectory and blue shaded regions show the 90\% prediction intervals. Solid grey lines are sample trajectories from the test set.}
    \label{fig:nodec_cp}
\end{figure}
When applying neural control models to real-world systems, it is essential to account for uncertainty due to noise. In this context, conformal prediction~\cite{vovk2005algorithmic,shafer2008tutorial} provides a model-agnostic and computationally efficient approach for uncertainty quantification in machine learning. The core idea is to begin with a point-prediction method and define a nonconformity score that quantifies the distance between a new sample and one from the calibration set. The conformal-prediction algorithm then transforms these scores into prediction regions with guaranteed coverage.

We apply split conformal prediction to quantify uncertainty in the dynamical system~\eqref{eq:mpc_1} controlled using NODEC. In our simulations, we incorporate both uncertainty in the initial condition (modeled as Gaussian noise with a standard deviation of 0.02) and additive process noise (Gaussian with a standard deviation of 0.1). We simulate multiple realizations at each time step. These samples are drawn independently from the same distribution and are therefore exchangeable, which is a key assumption in conformal prediction~\cite{vovk2005algorithmic}.

We implement split conformal prediction using the \texttt{PUNCC} library~\cite{mendil2023puncc}. The key steps are as follows:

\begin{enumerate}
\item We first compute a predictor $x_{(i,k)}$ for each state element $i \in \{1,2\}$ and time step $k \in \{0,\dots,N_T\}$, defined as the mean over $M$ realizations.

\item Next, we compute nonconformity scores for each time step $k \in \{0,\dots,N_T\}$ and calibration trajectory $j \in \{1,\dots,\widetilde{M}\}$. Specifically, for each state element $i \in \{1,2\}$, the nonconformity scores are
\begin{equation}
    s^{(j)}_{(i,k)}\left(x_{(i,k)}, \tilde{x}^{(j)}_{(i,k)}\right) = \left| x_{(i,k)} - \tilde{x}^{(j)}_{(i,k)} \right|\,,
\end{equation}
where $x_{(i,k)}$ denotes the $i$-th element of the mean trajectory at time step $k$, and $\tilde{x}^{(j)}_{(i,k)}$ is the corresponding element of the $j$-th noisy calibration trajectory.

\item For a new trajectory element $\tilde{x}_{(i,k)}^{(\widetilde{M}+1)}$, we compute the conformal prediction interval from the nonconformity scores as
\begin{equation}
    C\left(\tilde{x}_{(i,k)}^{(\widetilde{M}+1)}\right)=\left[C_-\left(\tilde{x}_{(i,k)}^{(\widetilde{M}+1)}\right),C_+\left(\tilde{x}_{(i,k)}^{(\widetilde{M}+1)}\right)\right]\,.
\end{equation}
The corresponding interval bounds are
\begin{align}
\begin{split}
    C_\pm\left(\tilde{x}_{(i,k)}^{(\widetilde{M}+1)}\right) = x_{(i,k)} \pm Q_{1-\alpha}& \Bigg( \sum_{j=1}^{\widetilde{M}} \frac{1}{\widetilde{M}+1} \cdot \delta_{s^{(j)}_{(i,k)}} \\
    & + \frac{1}{\widetilde{M}+1} \cdot \delta_{+\infty} \Bigg)\,,
\end{split}
\label{eq:cp_quantile}
\end{align}
where $Q_{1-\alpha}$ is the $(1 - \alpha)$-quantile of the weighted empirical distribution, and $\delta_x$ is a point mass at $x$.

The probability that the new trajectory element $\tilde{x}{(i,k)}^{(\widetilde{M}+1)}$ lies within the prediction interval satisfies
\begin{equation}
    \Pr\left(\tilde{x}_{(i,k)}^{(\widetilde{M}+1)} \in C\left(\tilde{x}_{(i,k)}^{(\widetilde{M}+1)}\right)\right) \geq 1 - \alpha\,.
\end{equation}
(See Theorem 1 in \cite{barber2023conformal} and Proposition 1 in \cite{DBLP:conf/ecml/PapadopoulosPVG02}.)
\end{enumerate}

Figure~\ref{fig:nodec_cp} shows the mean predicted trajectory (solid blue line) based on $M=100$ realizations, as well as the 90\% conformal prediction intervals (blue shaded regions), computed using $\widetilde{M} = 100$ calibration trajectories. Sample test trajectories are shown as solid grey lines. Most test trajectories lie well within the predicted intervals, demonstrating the effectiveness of the conformal prediction approach in quantifying uncertainty when applying NODEC in noisy environments.

If multiple process realizations are not available in a given problem (\ie, if only a single time series is observed), the exchangeability assumption underlying standard conformal prediction becomes particularly restrictive, as time-series data are typically temporally correlated. To address this limitation, researchers have proposed a class of methods based on weighted quantiles~\cite{barber2023conformal}, which relax the exchangeability requirement by assigning lower weights to observations that are further back in time.

Instead of assigning uniform weights $1/(\widetilde{M}+1)$ in Eq.~\eqref{eq:cp_quantile}, one can, for instance, use geometrically decaying weights such as
\begin{equation}
    w_{(i,k)} = \rho^{N_T - k}
\end{equation}
with $\rho\in(0,1)$ and normalize them according to
\begin{equation}
    \tilde{w}_{(i,k)} = \frac{w_{(i,k)}}{W}\,,\quad \text{where}~ W=\sum_{k=0}^{N_T} w_{(i,k)}+1\,.
\end{equation}
The weight of a future (unseen) observation is $\tilde{w}_{N_T+1} = W^{-1}$. This weighting scheme is consistent with approaches in~\cite{barber2023conformal, mendil2023puncc, chee2023uncertainty}. These weights are then used to compute a prediction interval for a future time step.
 
In addition to the example that we considered in this section, conformal prediction has been integrated into MPC frameworks where neural networks are used to generate prediction regions~\cite{lindemann2023safe}. It has also been employed in MPC tasks where neural networks model unknown system dynamics~\cite{chee2022knode, chee2023uncertainty}. In the context of nonlinear models of biological systems, conformal prediction has enabled uncertainty quantification up to two orders of magnitude faster than a Bayesian method, under the assumption of homoscedastic noise (\ie, constant variance of measurement errors relative to the signal across the time horizon)~\cite{portela2025conformal}. Related work has also applied conformal prediction to provide uncertainty quantification for neural surrogate models of partial differential equations~\cite{gopakumar2024uncertainty}.
\section{Conclusions}
\label{sec:conclusions}
\renewcommand{\arraystretch}{1.2}
\begin{table*}[b]
\centering
\caption{Relevant code repositories related to this work.}
\begin{tabular}{@{}l p{10cm}@{}}
\toprule
Topic & Repository link \\
\midrule
Application of NNC to two agent-based models & \url{https://gitlab.com/ComputationalScience/abm-control} \\
Application of NNC to inventory dynamics & \url{https://gitlab.com/ComputationalScience/idinn} \\
Application of NODEC to several dynamical systems & \url{https://github.com/asikist/nnc} \\
Additional NODEC examples & \url{https://gitlab.com/ComputationalScience/near-optimal-control} \\
Additional examples presented in this paper & \url{https://gitlab.com/ComputationalScience/neural-control} \\
Application of NODEC to microbiome dynamics & \url{https://github.com/danielriosgarza/AiSchool/tree/main/content/LucasBottcher} \\ 
Conformal prediction & \url{https://github.com/deel-ai/puncc} \\
\bottomrule
\end{tabular}
\label{tab:repo-links}
\end{table*}
As deep-learning and automatic-differentiation frameworks continue to improve, their application to control and optimization problems is becoming increasingly practical.

In this paper, we first reviewed selected neural-control approaches for discrete- and continuous-time dynamical systems, in both deterministic and stochastic settings, complemented by new examples that illustrate key concepts and highlight possible extensions for future research. We focused on applications across various domains such as biology, engineering, physics, and medicine. For continuous-time dynamical systems, neural ordinary differential equations (neural ODEs) provide a flexible framework for control-input parameterization. For discrete-time systems, we showed how custom neural control-input parameterizations can be implemented and optimized via automatic differentiation.

We then compared the differing optimization paradigms underlying model predictive control (MPC) and neural ODE control (NODEC), and discussed related approaches that incorporate neural networks into MPC. While NODEC typically relies on a fixed integration scheme during training, further research is needed to understand how changes in time step size or solver configuration affect generalization and performance at deployment~\cite{mohan2024you,nair2025understanding}. Finally, we integrated conformal prediction into noisy, neural-controlled dynamical systems to generate prediction intervals with guaranteed statistical coverage. We summarize code repositories associated with our work and related examples in Table~\ref{tab:repo-links}.

In several of the systems we studied, such as the block-move example, oscillator control, and the nonlinear system analyzed in the NODEC-MPC comparison, we observed that minimizing the distance to the terminal state can also implicitly reduce the running cost. Recent work~\cite{miao2025opt} presents an example demonstrating improved convergence of neural controllers that leverage a control Lyapunov function, compared to those focused solely on terminal-cost minimization.

We also highlighted the utility of a straight-through estimator for obtaining integer-valued control inputs in both the predator-prey agent-based model and the inventory dynamics example. Similar techniques have been applied in other domains, including recommender system design~\cite{asikis2023multi} and the calibration of financial agent-based models~\cite{dyer2023}. However, broader adoption of this approach will likely require advances in differentiating through more complex dynamics, particularly those involving discrete or stochastic elements (\eg, personalized medical digital twins~\cite{knapp2025personalizing}). In this context, stochastic automatic differentiation techniques show considerable promise~\cite{DBLP:conf/nips/AryaSSR22,DBLP:conf/atal/ChopraRSQKPR23}. Connections to differentiable variants of the Gillespie algorithm~\cite{rijal2025differentiable} also warrant further exploration. Alternatively, gradient-free methods such as ensemble Kalman inversion~\cite{bottcher2023gradient} may provide a viable path forward, especially in settings where gradients are ill-defined or intractable.
\begin{auth}
    LB is the sole author of this manuscript.
\end{auth}
\begin{funding}
    This work was supported by hessian.AI.
\end{funding}
\begin{acknowledgements}
 I am grateful to Thomas Asikis, Nino Antulov-Fantulin, Ioannis Fragkos, Jiawei Li, Luis L.\ Fonseca, Marcos Matabuena, and Reinhard C.\ Laubenbacher for valuable discussions. I also thank Daniel Garza for organizing the \href{https://ai-microbiome-school.onrender.com/}{School of Artificial Intelligence Applied to Microbiomes} at AgroParisTech.
 
\end{acknowledgements}
\renewcommand{\refname}{References}
\bibliographystyle{unsrt}
\bibliography{refs}   

\begin{thebibliography}{100}

\bibitem{rosenblueth1943behavior}
Arturo Rosenblueth, Norbert Wiener, and Julian Bigelow.
\newblock Behavior, purpose and teleology.
\newblock {\em Philosophy of Science}, 10(1):18--24, 1943.

\bibitem{wiener2019cybernetics}
Norbert Wiener.
\newblock {\em Cybernetics or control and communication in the animal and the
  machine}.
\newblock MIT Press, Boston, MA, USA, 1948.

\bibitem{thorndike1932fundamentals}
Edward Thorndike.
\newblock {\em Fundamentals of learning}.
\newblock Teachers College, Columbia University, New York City, NY, USA, 1932.

\bibitem{churchland1992computational}
Patricia~S. Churchland and Terrence~J. Sejnowski.
\newblock {\em The computational brain}.
\newblock MIT Press, Cambridge, MA, USA, 1992.

\bibitem{rumelhart1986pdp}
David~E. Rumelhart and James~L. McClelland.
\newblock {\em Parallel distributed processing, Volume 1: Explorations in the
  microstructure of cognition: Foundations}.
\newblock MIT Press, Cambridge, MA, USA, 1986.

\bibitem{rumelhart1986pdp2}
David~E. Rumelhart and James~L. McClelland.
\newblock {\em Parallel distributed processing, Volume 2: Explorations in the
  microstructure of cognition: Psychological and biological models}.
\newblock MIT Press, Cambridge, MA, USA, 1986.

\bibitem{miller1995neural}
W~Thomas Miller, Paul~J Werbos, and Richard~S Sutton.
\newblock {\em Neural networks for control}.
\newblock MIT Press, Cambridge, MA, 1995.

\bibitem{paszke2017automatic}
Adam Paszke, Sam Gross, Soumith Chintala, Gregory Chanan, Edward Yang, Zachary
  DeVito, Zeming Lin, Alban Desmaison, Luca Antiga, and Adam Lerer.
\newblock Automatic differentiation in {PyTorch}.
\newblock {\em NeurIPS Workshop on Autodiff}, 2017.

\bibitem{jax2018github}
James Bradbury, Roy Frostig, Peter Hawkins, Matthew~James Johnson, Chris Leary,
  Dougal Maclaurin, George Necula, Adam Paszke, Jake VanderPlas, Skye
  Wanderman-Milne, and Qiao Zhang.
\newblock {JAX}: composable transformations of {Python}+{NumPy} programs, 2018.

\bibitem{lutter2019deep}
Michael Lutter, Christian Ritter, and Jan Peters.
\newblock Deep {L}agrangian networks: Using physics as model prior for deep
  learning.
\newblock {\em arXiv preprint arXiv:1907.04490}, 2019.

\bibitem{zhong2019symplectic}
Yaofeng~Desmond Zhong, Biswadip Dey, and Amit Chakraborty.
\newblock {S}ymplectic {ODE-Net}: {L}earning {H}amiltonian dynamics with
  control.
\newblock {\em arXiv preprint arXiv:1909.12077}, 2019.

\bibitem{jin2019pontryagin}
Wanxin Jin, Zhaoran Wang, Zhuoran Yang, and Shaoshuai Mou.
\newblock Pontryagin differentiable programming: An end-to-end learning and
  control framework.
\newblock In Hugo Larochelle, Marc'Aurelio Ranzato, Raia Hadsell,
  Maria{-}Florina Balcan, and Hsuan{-}Tien Lin, editors, {\em Advances in
  Neural Information Processing Systems 33: Annual Conference on Neural
  Information Processing Systems 2020, NeurIPS 2020, December 6-12, 2020,
  virtual}, 2020.

\bibitem{asikis2022neural}
Thomas Asikis, Lucas B{\"o}ttcher, and Nino Antulov-Fantulin.
\newblock Neural ordinary differential equation control of dynamics on graphs.
\newblock {\em Physical Review Research}, 4(1):013221, 2022.

\bibitem{bottcher2022near}
Lucas B{\"o}ttcher and Thomas Asikis.
\newblock Near-optimal control of dynamical systems with neural ordinary
  differential equations.
\newblock {\em Machine Learning: Science and Technology}, 3(4):045004, 2022.

\bibitem{bottcher2022ai}
Lucas B{\"o}ttcher, Nino Antulov-Fantulin, and Thomas Asikis.
\newblock {AI} {P}ontryagin or how artificial neural networks learn to control
  dynamical systems.
\newblock {\em Nature Communications}, 13(1):333, 2022.

\bibitem{chee2022knode}
Kong~Yao Chee, Tom~Z Jiahao, and M~Ani Hsieh.
\newblock {KNODE}-{MPC}: A knowledge-based data-driven predictive control
  framework for aerial robots.
\newblock {\em IEEE Robotics and Automation Letters}, 7(2):2819--2826, 2022.

\bibitem{bottcher2023gradient}
Lucas B{\"o}ttcher.
\newblock Gradient-free training of neural {ODE}s for system identification and
  control using ensemble {K}alman inversion.
\newblock In {\em ICML Workshop on New Frontiers in Learning, Control, and
  Dynamical Systems, Honolulu, HI, USA, 2023}, 2023.

\bibitem{mowlavi2023optimal}
Saviz Mowlavi and Saleh Nabi.
\newblock Optimal control of {PDE}s using physics-informed neural networks.
\newblock {\em Journal of Computational Physics}, 473:111731, 2023.

\bibitem{bottcher2023control}
Lucas B{\"o}ttcher, Thomas Asikis, and Ioannis Fragkos.
\newblock Control of dual-sourcing inventory systems using recurrent neural
  networks.
\newblock {\em INFORMS Journal on Computing}, 35(6):1308--1328, 2023.

\bibitem{nghiem2023physics}
Truong~X Nghiem, J{\'a}n Drgo{\v{n}}a, Colin Jones, Zoltan Nagy, Roland Schwan,
  Biswadip Dey, Ankush Chakrabarty, Stefano Di~Cairano, Joel~A Paulson, Andrea
  Carron, et~al.
\newblock Physics-informed machine learning for modeling and control of
  dynamical systems.
\newblock In {\em 2023 American Control Conference (ACC)}, pages 3735--3750,
  2023.

\bibitem{chee2023uncertainty}
Kong~Yao Chee, M~Ani Hsieh, and George~J Pappas.
\newblock Uncertainty quantification for learning-based {MPC} using weighted
  conformal prediction.
\newblock In {\em 2023 62nd IEEE Conference on Decision and Control (CDC)},
  pages 342--349, 2023.

\bibitem{mou2024model}
Muyun Mou, Yu~Guo, Fanming Luo, Yang Yu, and Jiang Zhang.
\newblock Model predictive complex system control from observational and
  interventional data.
\newblock {\em Chaos: An Interdisciplinary Journal of Nonlinear Science},
  34(9), 2024.

\bibitem{chen2024accelerated}
Song Chen, Jiaxu Liu, Pengkai Wang, Chao Xu, Shengze Cai, and Jian Chu.
\newblock Accelerated optimization in deep learning with a
  proportional-integral-derivative controller.
\newblock {\em Nature Communications}, 15(1):10263, 2024.

\bibitem{DELALEAU2025229}
Emmanuel Delaleau, Cédric Join, and Michel Fliess.
\newblock Synchronization of kuramoto oscillators via {HEOL}, and a discussion
  on {AI}.
\newblock {\em IFAC-PapersOnLine}, 59(1):229--234, 2025.
\newblock 11th Vienna International Conference on Mathematical Modelling
  MATHMOD 2025.

\bibitem{bottcher2025control}
Lucas B{\"o}ttcher, Luis~L Fonseca, and Reinhard~C Laubenbacher.
\newblock Control of medical digital twins with artificial neural networks.
\newblock {\em Philosophical Transactions A}, 383(2292):20240228, 2025.

\bibitem{wang2025pidnodes}
Pengkai Wang, Song Chen, Jiaxu Liu, Shengze Cai, and Chao Xu.
\newblock {PIDNODE}s: Neural ordinary differential equations inspired by a
  proportional--integral--derivative controller.
\newblock {\em Neurocomputing}, 614:128769, 2025.

\bibitem{raghu2017deep}
Aniruddh Raghu, Matthieu Komorowski, Imran Ahmed, Leo Celi, Peter Szolovits,
  and Marzyeh Ghassemi.
\newblock Deep reinforcement learning for sepsis treatment.
\newblock {\em arXiv preprint arXiv:1711.09602}, 2017.

\bibitem{wen2019online}
Yue Wen, Jennie Si, Andrea Brandt, Xiang Gao, and He~Helen Huang.
\newblock Online reinforcement learning control for the personalization of a
  robotic knee prosthesis.
\newblock {\em IEEE Transactions on Cybernetics}, 50(6):2346--2356, 2019.

\bibitem{steffen2025deep}
Sebastian Steffen and Mark Cannon.
\newblock Deep learning model predictive control for deep brain stimulation in
  {P}arkinson's disease.
\newblock {\em arXiv preprint arXiv:2504.00618}, 2025.

\bibitem{deng2023alibaba}
Yuming Deng, Xinhui Zhang, Tong Wang, Lin Wang, Yidong Zhang, Xiaoqing Wang,
  Su~Zhao, Yunwei Qi, Guangyao Yang, and Xuezheng Peng.
\newblock Alibaba realizes millions in cost savings through integrated demand
  forecasting, inventory management, price optimization, and product
  recommendations.
\newblock {\em INFORMS Journal on Applied Analytics}, 53(1):32--46, 2023.

\bibitem{Degrave2022}
Jonas Degrave, Federico Felici, Jonas Buchli, Michael Neunert, Brendan Tracey,
  Francesco Carpanese, Timo Ewalds, Roland Hafner, Abbas Abdolmaleki, Diego
  de~las Casas, Craig Donner, Leslie Fritz, Cristian Galperti, Andrea Huber,
  James Keeling, Maria Tsimpoukelli, Jackie Kay, Antoine Merle, Jean-Marc
  Moret, Seb Noury, Federico Pesamosca, David Pfau, Olivier Sauter, Cristian
  Sommariva, Stefano Coda, Basil Duval, Ambrogio Fasoli, Pushmeet Kohli, Koray
  Kavukcuoglu, Demis Hassabis, and Martin Riedmiller.
\newblock Magnetic control of tokamak plasmas through deep reinforcement
  learning.
\newblock {\em Nature}, 602(7897):414--419, 2022.

\bibitem{schmidhuber2015deep}
J{\"u}rgen Schmidhuber.
\newblock Deep learning in neural networks: An overview.
\newblock {\em Neural Networks}, 61:85--117, 2015.

\bibitem{ivakhnenko1971polynomial}
Alexey~Grigorevich Ivakhnenko.
\newblock Polynomial theory of complex systems.
\newblock {\em IEEE Transactions on Systems, Man, and Cybernetics},
  (4):364--378, 1971.

\bibitem{lewis2020neural}
FW~Lewis, Suresh Jagannathan, and Aydin Yesildirak.
\newblock {\em Neural network control of robot manipulators and non-linear
  systems}.
\newblock CRC Press, Boca Raton, FL, 2020.

\bibitem{chen2018neural}
Ricky T.~Q. Chen, Yulia Rubanova, Jesse Bettencourt, and David Duvenaud.
\newblock Neural ordinary differential equations.
\newblock {\em Advances in Neural Information Processing Systems}, 2018.

\bibitem{wang1998runge}
Yi-Jen Wang and Chin-Teng Lin.
\newblock {R}unge-{K}utta neural network for identification of dynamical
  systems in high accuracy.
\newblock {\em IEEE Transactions on Neural Networks}, 9(2):294--307, 1998.

\bibitem{abu2005nearly}
Murad Abu-Khalaf and Frank~L Lewis.
\newblock Nearly optimal control laws for nonlinear systems with saturating
  actuators using a neural network {HJB} approach.
\newblock {\em Automatica}, 41(5):779--791, 2005.

\bibitem{PhysRevFluids.4.093902}
Guido Novati, L.~Mahadevan, and Petros Koumoutsakos.
\newblock Controlled gliding and perching through deep-reinforcement-learning.
\newblock {\em Physical Review Fluids}, 4:093902, 2019.

\bibitem{de2023deep}
Agostino De~Marco, Paolo~Maria D’Onza, and Sabato Manfredi.
\newblock A deep reinforcement learning control approach for high-performance
  aircraft.
\newblock {\em Nonlinear Dynamics}, 111(18):17037--17077, 2023.

\bibitem{gu2023optimal}
Zhiyang Gu, Chengli Fan, Dengxiu Yu, and Zhen Wang.
\newblock Optimal synchronized control of nonlinear coupled harmonic
  oscillators based on actor--critic reinforcement learning.
\newblock {\em Nonlinear Dynamics}, 111(22):21051--21064, 2023.

\bibitem{wang2025energy}
Xiaolong Wang, Jianfu Cao, Ye~Cao, and Feng Zou.
\newblock Energy-efficient trajectory planning for a class of industrial robots
  using parallel deep reinforcement learning.
\newblock {\em Nonlinear Dynamics}, 113(8):8491--8511, 2025.

\bibitem{mizutani2004two}
Eiji Mizutani and Stuart~E Dreyfus.
\newblock Two stochastic dynamic programming problems by model-free
  actor-critic recurrent-network learning in non-{M}arkovian settings.
\newblock In {\em 2004 IEEE International Joint Conference on Neural Networks
  (IEEE Cat. No. 04CH37541)}, volume~2, pages 1079--1084. IEEE, 2004.

\bibitem{DBLP:conf/nips/JinABJ18}
Chi Jin, Zeyuan Allen{-}Zhu, S{\'{e}}bastien Bubeck, and Michael~I. Jordan.
\newblock Is {Q}-learning provably efficient?
\newblock In Samy Bengio, Hanna~M. Wallach, Hugo Larochelle, Kristen Grauman,
  Nicol{\`{o}} Cesa{-}Bianchi, and Roman Garnett, editors, {\em Advances in
  Neural Information Processing Systems 31: Annual Conference on Neural
  Information Processing Systems 2018, NeurIPS 2018, December 3-8, 2018,
  Montr{\'{e}}al, Canada}, pages 4868--4878, 2018.

\bibitem{DBLP:conf/aaai/Yarats0KAPF21}
Denis Yarats, Amy Zhang, Ilya Kostrikov, Brandon Amos, Joelle Pineau, and Rob
  Fergus.
\newblock Improving sample efficiency in model-free reinforcement learning from
  images.
\newblock In {\em Thirty-Fifth {AAAI} Conference on Artificial Intelligence,
  {AAAI} 2021, Thirty-Third Conference on Innovative Applications of Artificial
  Intelligence, {IAAI} 2021, The Eleventh Symposium on Educational Advances in
  Artificial Intelligence, {EAAI} 2021, Virtual Event, February 2-9, 2021},
  pages 10674--10681. {AAAI} Press, 2021.

\bibitem{beverton1957dynamics}
R.~J.~H. Beverton and S.~J. Holt.
\newblock {\em On the Dynamics of Exploited Fish Populations}, volume~19 of
  {\em Fisheries Investigations Series II}.
\newblock Ministry of Agriculture, Fisheries and Food, London, UK, 1957.

\bibitem{baranov1918biological}
F.~I. Baranov.
\newblock On the question of the biological basis of fisheries.
\newblock {\em Izvestiya Otdella Rybolovstva i Nauchno-Promyslovykh
  Issledovanii}, 1:81--128, 1918.
\newblock In Russian.

\bibitem{kenchington2021baranov}
Trevor~J Kenchington.
\newblock Baranov's contributions to the {B}everton--{H}olt model.
\newblock {\em ICES Journal of Marine Science}, 78(6):2166--2172, 2021.

\bibitem{whittle_discrete_eco_models}
Andrew Whittle.
\newblock Discrete time mathematical models in ecology.
\newblock University of Tennessee, Department of Mathematics.

\bibitem{may1975nonlinear}
Robert~M May and Warren~J Leonard.
\newblock Nonlinear aspects of competition between three species.
\newblock {\em SIAM Journal on Applied Mathematics}, 29(2):243--253, 1975.

\bibitem{pekalski1998three}
Andrzej Pekalski and Dietrich Stauffer.
\newblock Three species {L}otka--{V}olterra model.
\newblock {\em International Journal of Modern Physics C}, 9(05):777--783,
  1998.

\bibitem{wilensky1997netlogo}
Uri Wilensky.
\newblock {NetLogo} {W}olf {S}heep {P}redation {M}odel.
\newblock {\em Center for Connected Learning and Computer-Based Modeling,
  Northwestern University}, 1997.

\bibitem{wilensky1999netlogo}
Uri Wilensky.
\newblock {NetLogo}.
\newblock \url{http://ccl.northwestern.edu/netlogo/}, 1999.

\bibitem{oremland2016computational}
Matthew Oremland, Kathryn~R Michels, Alexandra~M Bettina, Chris Lawrence, Borna
  Mehrad, and Reinhard Laubenbacher.
\newblock A computational model of invasive aspergillosis in the lung and the
  role of iron.
\newblock {\em BMC Systems Biology}, 10(1):1--14, 2016.

\bibitem{ribeiro2022multi}
Henrique~AL Ribeiro, Luis~Sordo Vieira, Yogesh Scindia, Bandita Adhikari,
  Matthew Wheeler, Adam Knapp, William Schroeder, Borna Mehrad, and Reinhard
  Laubenbacher.
\newblock Multi-scale mechanistic modelling of the host defence in invasive
  aspergillosis reveals leucocyte activation and iron acquisition as drivers of
  infection outcome.
\newblock {\em Journal of the Royal Society Interface}, 19(189):20210806, 2022.

\bibitem{faust2012microbial}
Karoline Faust and Jeroen Raes.
\newblock Microbial interactions: from networks to models.
\newblock {\em Nature Reviews Microbiology}, 10(8):538--550, 2012.

\bibitem{714215}
Sukhan Lee and Jun Park.
\newblock Dual-mode dynamics neural network ({D2NN}) for knapsack packing
  problem.
\newblock In {\em Proceedings of 1993 International Conference on Neural
  Networks (IJCNN-93-Nagoya, Japan)}, volume~3, pages 2425--2428, 1993.

\bibitem{asikis2023multi}
Thomas Asikis.
\newblock Towards recommendations for value sensitive sustainable consumption.
\newblock In {\em {N}eur{IPS} 2023 {W}orkshop on {T}ackling {C}limate {C}hange
  with {M}achine {L}earning: {B}lending {N}ew and {E}xisting {K}nowledge
  {S}ystems}, 2023.

\bibitem{dyer2023}
Joel Dyer, Arnau Quera{-}Bofarull, Ayush Chopra, J.~Doyne Farmer, Anisoara
  Calinescu, and Michael~J. Wooldridge.
\newblock Gradient-assisted calibration for financial agent-based models.
\newblock In {\em 4th {ACM} {I}nternational {C}onference on {AI} in {F}inance,
  {ICAIF} 2023, {B}rooklyn, {NY}, {USA}, {N}ovember 27-29, 2023}, pages
  288--296. {ACM}, 2023.

\bibitem{fonseca2025optimal}
Luis~L Fonseca, Lucas B{\"o}ttcher, Borna Mehrad, and Reinhard~C Laubenbacher.
\newblock Optimal control of agent-based models via surrogate modeling.
\newblock {\em PLOS Computational Biology}, 21(1):e1012138, 2025.

\bibitem{gijsbrechts2025ai}
Joren Gijsbrechts, Robert~N Boute, Jan~A Van~Mieghem, and Dennis Zhang.
\newblock {AI} in inventory management: The disruptive era of {DRL} and beyond.
\newblock {\em Available at SSRN}, 2025.

\bibitem{gijsbrechts2022can}
Joren Gijsbrechts, Robert~N Boute, Jan~A Van~Mieghem, and Dennis~J Zhang.
\newblock Can deep reinforcement learning improve inventory management?
  performance on lost sales, dual-sourcing, and multi-echelon problems.
\newblock {\em Manufacturing \& Service Operations Management},
  24(3):1349--1368, 2022.

\bibitem{barankin1961delivery}
Edward Barankin.
\newblock A delivery-lag inventory model with an emergency provision.
\newblock {\em Naval Research Logistics Quarterly}, 8:285--311, 1961.

\bibitem{fukuda1964optimal}
Yasunari Fukuda.
\newblock Optimal policies for the inventory problem with negotiable leadtime.
\newblock {\em Management Science}, 10(4):690--708, 1964.

\bibitem{xin2023dual}
Linwei Xin and Jan~A Van~Mieghem.
\newblock Dual-sourcing, dual-mode dynamic stochastic inventory models.
\newblock In {\em {R}esearch {H}andbook on {I}nventory {M}anagement}, pages
  165--190. Edward Elgar Publishing, Cheltenham, UK, 2023.

\bibitem{arrow1951optimal}
Kenneth~J. Arrow, Theodore Harris, and Jacob Marschak.
\newblock Optimal inventory policy.
\newblock {\em Econometrica}, 19(3):250--272, 1951.

\bibitem{scarf1958inventory}
Herbert Scarf and Samuel Karlin.
\newblock Inventory models of the {A}rrow-{H}arris-{M}arschak type with time
  lag.
\newblock In Kenneth~J. Arrow, Samuel Karlin, and Herbert~E. Scarf, editors,
  {\em Studies in the Mathematical Theory of Inventory and Production}.
  Stanford University Press, Stanford, CA, 1958.

\bibitem{DBLP:conf/acssc/DouglasY18}
Scott~C. Douglas and Jiutian Yu.
\newblock Why {RELU} units sometimes die: Analysis of single-unit error
  backpropagation in neural networks.
\newblock In Michael~B. Matthews, editor, {\em 52nd Asilomar Conference on
  Signals, Systems, and Computers, {ACSSC} 2018, Pacific Grove, CA, USA,
  October 28-31, 2018}, pages 864--868. {IEEE}, 2018.

\bibitem{barron2017continuously}
Jonathan~T Barron.
\newblock Continuously differentiable exponential linear units.
\newblock {\em arXiv preprint arXiv:1704.07483}, 2017.

\bibitem{hua2015structural}
Zhongsheng Hua, Yimin Yu, Wei Zhang, and Xiaoyan Xu.
\newblock Structural properties of the optimal policy for dual-sourcing systems
  with general lead times.
\newblock {\em IIE Transactions}, 47(8):841--850, 2015.

\bibitem{sun2019robust}
Jiankun Sun and Jan~A Van~Mieghem.
\newblock Robust dual sourcing inventory management: Optimality of capped dual
  index policies and smoothing.
\newblock {\em Manufacturing \& Service Operations Management}, 21(4):912--931,
  2019.

\bibitem{gong2006pseudospectral}
Qi~Gong, Wei Kang, and I~Michael Ross.
\newblock A pseudospectral method for the optimal control of constrained
  feedback linearizable systems.
\newblock {\em IEEE Transactions on Automatic Control}, 51(7):1115--1129, 2006.

\bibitem{he2015delving}
Kaiming He, Xiangyu Zhang, Shaoqing Ren, and Jian Sun.
\newblock {Delving deep into rectifiers: Surpassing human-level performance on
  ImageNet classification}.
\newblock In {\em Proceedings of the IEEE International Conference on Computer
  Vision}, pages 1026--1034, 2015.

\bibitem{buffie2012profound}
Charlie~G Buffie, Irene Jarchum, Michele Equinda, Lauren Lipuma, Asia Gobourne,
  Agnes Viale, Carles Ubeda, Joao Xavier, and Eric~G Pamer.
\newblock Profound alterations of intestinal microbiota following a single dose
  of clindamycin results in sustained susceptibility to \textit{{C}lostridium
  difficile}-induced colitis.
\newblock {\em Infection and Immunity}, 80(1):62--73, 2012.

\bibitem{stein2013ecological}
Richard~R Stein, Vanni Bucci, Nora~C Toussaint, Charlie~G Buffie, Gunnar
  R{\"a}tsch, Eric~G Pamer, Chris Sander, and Joao~B Xavier.
\newblock Ecological modeling from time-series inference: insight into dynamics
  and stability of intestinal microbiota.
\newblock {\em PLOS Computational Biology}, 9(12):e1003388, 2013.

\bibitem{jones2018silico}
Eric~W Jones and Jean~M Carlson.
\newblock \textit{In silico} analysis of antibiotic-induced
  \textit{{C}lostridium difficile} infection: Remediation techniques and
  biological adaptations.
\newblock {\em PLOS Computational Biology}, 14(2):e1006001, 2018.

\bibitem{jones2020navigation}
E.~W. Jones, P.~Shankin-Clarke, and J.~M. Carlson.
\newblock Navigation and control of outcomes in a generalized
  {L}otka--{V}olterra model of the microbiome.
\newblock In J.~Kotas, editor, {\em Advances in Nonlinear Biological Systems:
  Modeling and Optimal Control}, volume~11 of {\em AIMS Series on Applied
  Mathematics}, pages 97--120. American Institute of Mathematical Sciences,
  Springfield, MO, USA, 2020.

\bibitem{bonnard2024geometric}
Bernard Bonnard, J{\'e}r{\'e}my Rouot, and Cristiana~J Silva.
\newblock Geometric optimal control of the generalized {L}otka--{V}olterra
  model of the intestinal microbiome.
\newblock {\em Optimal Control Applications and Methods}, 45(2):544--574, 2024.

\bibitem{mcfarland2002breaking}
Lynne~V McFarland, Gary~W Elmer, and Christina~M Surawicz.
\newblock Breaking the cycle: treatment strategies for 163 cases of
  recurrentclostridium difficiledisease.
\newblock {\em Official Journal of the American College of Gastroenterology|
  ACG}, 97(7):1769--1775, 2002.

\bibitem{kuramoto1975self}
Yoshiki Kuramoto.
\newblock Self-entrainment of a population of coupled non-linear oscillators.
\newblock In {\em International Symposium on Mathematical Problems in
  Theoretical Physics}, pages 420--422. Springer, 1975.

\bibitem{ha2016emergence}
Seung-Yeal Ha, Hwa~Kil Kim, and Sang~Woo Ryoo.
\newblock Emergence of phase-locked states for the {K}uramoto model in a large
  coupling regime.
\newblock {\em Communications in Mathematical Sciences}, 14(4):1073--1091,
  2016.

\bibitem{biccari2020stochastic}
Umberto Biccari and Enrique Zuazua.
\newblock {A stochastic approach to the synchronization of coupled
  oscillators}.
\newblock {\em Frontiers in Energy Research}, 8(115), 2020.

\bibitem{dorfler2013synchronization}
Florian D{\"o}rfler, Michael Chertkov, and Francesco Bullo.
\newblock Synchronization in complex oscillator networks and smart grids.
\newblock {\em Proceedings of the National Academy of Sciences USA},
  110(6):2005--2010, 2013.

\bibitem{liu2011controllability}
Yang-Yu Liu, Jean-Jacques Slotine, and Albert-L{\'a}szl{\'o} Barab{\'a}si.
\newblock Controllability of complex networks.
\newblock {\em Nature}, 473(7346):167--173, 2011.

\bibitem{paluch2024hardware}
Marcin Paluch, Florian Bolli, Xiang Deng, Antonio~Rios Navarro, Chang Gao, and
  Tobi Delbruck.
\newblock Hardware neural control of {C}art{P}ole and {F1TENTH} race car.
\newblock {\em arXiv preprint arXiv:2407.08681}, 2024.

\bibitem{saint1991neural}
Jean Saint-Donat, Naveen Bhat, and Thomas~J McAvoy.
\newblock Neural net based model predictive control.
\newblock {\em International Journal of Control}, 54(6):1453--1468, 1991.

\bibitem{draeger1995model}
Andreas Draeger, Sebastian Engell, and Horst Ranke.
\newblock Model predictive control using neural networks.
\newblock {\em IEEE Control Systems Magazine}, 15(5):61--66, 1995.

\bibitem{DBLP:conf/icml/AmosXK17}
Brandon Amos, Lei Xu, and J.~Zico Kolter.
\newblock Input convex neural networks.
\newblock In Doina Precup and Yee~Whye Teh, editors, {\em Proceedings of the
  34th International Conference on Machine Learning, {ICML} 2017, Sydney, NSW,
  Australia, 6-11 August 2017}, volume~70 of {\em Proceedings of Machine
  Learning Research}, pages 146--155. {PMLR}, 2017.

\bibitem{DBLP:conf/l4dc/BunningSABHL21}
Felix B{\"{u}}nning, Adrian Schalbetter, Ahmed Aboudonia, Mathias~Hudoba
  de~Badyn, Philipp Heer, and John Lygeros.
\newblock Input convex neural networks for building {MPC}.
\newblock In Ali Jadbabaie, John Lygeros, George~J. Pappas, Pablo~A. Parrilo,
  Benjamin Recht, Claire~J. Tomlin, and Melanie~N. Zeilinger, editors, {\em
  Proceedings of the 3rd Annual Conference on Learning for Dynamics and
  Control, {L4DC} 2021, 7-8 June 2021, Virtual Event, Switzerland}, volume 144
  of {\em Proceedings of Machine Learning Research}, pages 251--262. {PMLR},
  2021.

\bibitem{kaiser2018sparse}
Eurika Kaiser, J~Nathan Kutz, and Steven~L Brunton.
\newblock Sparse identification of nonlinear dynamics for model predictive
  control in the low-data limit.
\newblock {\em Proceedings of the Royal Society A}, 474(2219):20180335, 2018.

\bibitem{vovk2005algorithmic}
Vladimir Vovk, Alexander Gammerman, and Glenn Shafer.
\newblock {\em Algorithmic Learning in a Random World}, volume~29.
\newblock Springer, New York, NY, USA, 2005.

\bibitem{shafer2008tutorial}
Glenn Shafer and Vladimir Vovk.
\newblock A tutorial on conformal prediction.
\newblock {\em Journal of Machine Learning Research}, 9(3), 2008.

\bibitem{mendil2023puncc}
Mouhcine Mendil, Luca Mossina, and David Vigouroux.
\newblock {PUNCC:} a python library for predictive uncertainty calibration and
  conformalization.
\newblock In Harris Papadopoulos, Khuong~An Nguyen, Henrik Bostr{\"{o}}m, and
  Lars Carlsson, editors, {\em Conformal and Probabilistic Prediction with
  Applications, 13-15 September 2023, Limassol, Cyprus}, volume 204 of {\em
  Proceedings of Machine Learning Research}, pages 582--601. {PMLR}, 2023.

\bibitem{barber2023conformal}
Rina~Foygel Barber, Emmanuel~J Candes, Aaditya Ramdas, and Ryan~J Tibshirani.
\newblock Conformal prediction beyond exchangeability.
\newblock {\em The Annals of Statistics}, 51(2):816--845, 2023.

\bibitem{DBLP:conf/ecml/PapadopoulosPVG02}
Harris Papadopoulos, Kostas Proedrou, Volodya Vovk, and Alex Gammerman.
\newblock Inductive confidence machines for regression.
\newblock In Tapio Elomaa, Heikki Mannila, and Hannu Toivonen, editors, {\em
  Machine Learning: {ECML} 2002, 13th European Conference on Machine Learning,
  Helsinki, Finland, August 19-23, 2002, Proceedings}, volume 2430 of {\em
  Lecture Notes in Computer Science}, pages 345--356. Springer, 2002.

\bibitem{lindemann2023safe}
Lars Lindemann, Matthew Cleaveland, Gihyun Shim, and George~J Pappas.
\newblock Safe planning in dynamic environments using conformal prediction.
\newblock {\em IEEE Robotics and Automation Letters}, 8(8):5116--5123, 2023.

\bibitem{portela2025conformal}
Alberto Portela, Julio~R Banga, and Marcos Matabuena.
\newblock Conformal prediction for uncertainty quantification in dynamic
  biological systems.
\newblock {\em PLOS Computational Biology}, 21(5):e1013098, 2025.

\bibitem{gopakumar2024uncertainty}
Vignesh Gopakumar, Ander Gray, Joel Oskarsson, Lorenzo Zanisi, Stanislas
  Pamela, Daniel Giles, Matt Kusner, and Marc~Peter Deisenroth.
\newblock Uncertainty quantification of surrogate models using conformal
  prediction.
\newblock {\em arXiv preprint arXiv:2408.09881}, 2024.

\bibitem{mohan2024you}
Arvind Mohan, Ashesh Chattopadhyay, and Jonah Miller.
\newblock What you see is not what you get: Neural partial differential
  equations and the illusion of learning.
\newblock {\em arXiv preprint arXiv:2411.15101}, 2024.

\bibitem{nair2025understanding}
Ashish~S Nair, Shivam Barwey, Pinaki Pal, Jonathan~F MacArt, Troy Arcomano, and
  Romit Maulik.
\newblock Understanding latent timescales in neural ordinary differential
  equation models of advection-dominated dynamical systems.
\newblock {\em Physica D: Nonlinear Phenomena}, 476:134650, 2025.

\bibitem{miao2025opt}
Keyan Miao, Liqun Zhao, Han Wang, Konstantinos Gatsis, and Antonis
  Papachristodoulou.
\newblock Opt-{ODEN}et: A neural {ODE} framework with differentiable {QP}
  layers for safe and stable control design (longer version).
\newblock {\em arXiv preprint arXiv:2504.17139}, 2025.

\bibitem{knapp2025personalizing}
Adam Knapp, Daniel~A. Cruz, Borna Mehrad, and Reinhard~C. Laubenbacher.
\newblock Personalizing computational models to construct medical digital
  twins.
\newblock {\em Journal of the Royal Society Interface}, 22(20250055), 2025.

\bibitem{DBLP:conf/nips/AryaSSR22}
Gaurav Arya, Moritz Schauer, Frank Sch{\"{a}}fer, and Christopher Rackauckas.
\newblock Automatic differentiation of programs with discrete randomness.
\newblock In Sanmi Koyejo, S.~Mohamed, A.~Agarwal, Danielle Belgrave, K.~Cho,
  and A.~Oh, editors, {\em Advances in Neural Information Processing Systems
  35: Annual Conference on Neural Information Processing Systems 2022, NeurIPS
  2022, New Orleans, LA, USA, November 28 - December 9, 2022}, 2022.

\bibitem{DBLP:conf/atal/ChopraRSQKPR23}
Ayush Chopra, Alexander Rodr{\'{\i}}guez, Jayakumar Subramanian, Arnau
  Quera{-}Bofarull, Balaji Krishnamurthy, B.~Aditya Prakash, and Ramesh Raskar.
\newblock Differentiable agent-based epidemiology.
\newblock In Noa Agmon, Bo~An, Alessandro Ricci, and William Yeoh, editors,
  {\em Proceedings of the 2023 International Conference on Autonomous Agents
  and Multiagent Systems, {AAMAS} 2023, London, United Kingdom, 29 May 2023 - 2
  June 2023}, pages 1848--1857. {ACM}, 2023.

\bibitem{rijal2025differentiable}
Krishna Rijal and Pankaj Mehta.
\newblock A differentiable {G}illespie algorithm for simulating chemical
  kinetics, parameter estimation, and designing synthetic biological circuits.
\newblock {\em ELife}, 14:RP103877, 2025.

\end{thebibliography}

\end{document}